\documentclass{aa}  

\usepackage{graphicx}
\usepackage{txfonts}

\begin{document} 

   \title{Comparison of the performance of the Shack-Hartmann and Pyramid wavefront sensors with a laser guide star for 40 m telescopes.}
   \titlerunning{Comparison of the performance of the SHWFS and PWFS with an LGS for 40 m telescopes}

   \author{F. Oyarzún
          \inst{1}
          C. Heritier \inst{3,1}
          V. Chambouleyron \inst{2}
          T. Fusco \inst{3,1}
          P. Rouquette \inst{1}
          \and
          B. Neichel \inst{1}
          }

   \institute{Aix Marseille Univ, CNRS, CNES, LAM, Marseille, France\\
              \email{francisco.oyarzun@lam.fr}
         \and
            University of California Santa Cruz, 1156 High St, Santa Cruz, USA\\
         \and
             DOTA, ONERA, Université Paris Saclay, F-91123 Palaiseau, France
             }

   \authorrunning{F. Oyarzún}
   \date{\today}

 
  \abstract
   {The new giant segmented mirror telescopes will use laser guide stars (LGS) for their adaptive optics (AO) systems. Two options to use as wavefront sensors (WFS) are the Shack-Hartmann wavefront sensor (SHWFS) and the pyramid wavefront sensor (PWFS).}
   {In this paper, we compare the noise performance of the PWFS and the SHWFS. We aim to find which of the two WFS is the best to use in a single or tomographic configuration.}
   {To compute the noise performance we extended a noise model developed for the PWFS to be used with the SHWFS. To do this, we expressed the centroiding algorithm of the SHWFS as a matrix-vector multiplication, which allowed us to use the statistics of noise to compute its propagation through the AO loop. We validated the noise model with end-to-end simulations for telescopes of 8 and 16 m in diameter.}
   {For an AO system with only one WFS, we found that, given the same number of subapertures, the PWFS outperforms the SHWFS. For a 40 m telescope, the limiting magnitude of the PWFS is around 1 magnitude higher than the SHWFS. When using multiple WFS and a Generalized least squares estimator to combine the signal, our model predicts that in a tomographic system, the SHWFS performs better than the PWFS having a limiting magnitude 0.3 magnitudes higher. If using sub-electron RON detectors for the PWFS, then the performances are almost identical between the two WFSs}
   {We conclude that when using a single WFS with LGS, the PWFS is a better alternative than the SH. However, for a tomographic system, either would have almost the same performance.}

    
   \maketitle
%

\section{Introduction}

The new Giant Segmented Mirror Telescopes (GSMT) will use adaptive optics (AO) systems to deal with the degradation in angular resolution given by atmospheric effects. With AO, it would be possible to obtain angular resolutions close to the theoretical limit, giving exciting opportunities to study small, distant objects. With the help of coronagraphic instruments, it would offer the capabilities of obtaining direct images of more exoplanets than ever before, and be able to study their atmospheric composition \citep{2016SPIE.9908E..1ZD}. As the GSMTs gather up to ten times more light than their predecessors, they will allow us to study fainter objects, unlocking new insights into the early universe and its formation \citep{2007Msngr.127...11G}.

The AO systems require the use of a guide object. The GSMTs will use Laser Guide Stars (LGS; \citealp{2016SPIE.9908E..1ZD, 2016SPIE.9908E..1XT, 2022SPIE12185E..14C}) to compensate for the lack of Natural Guide Stars (NGS) bright enough to have good sky coverage \citep{1991Natur.353..141P}. These LGS are generated by exciting sodium atoms present on a layer around 90 km above sea level \citep{1985A&A...152L..29F}. This layer is around 20 km thick and has an evolving sodium density profile \citep{2014A&A...565A.102P}. To generate the LGS a sodium laser is used. For this, dedicated laser launch telescopes are mounted at the side of the telescope that shoots beams of light calibrated at 589 nm. Due to beam divergence and atmospheric distortions, the beam has around a 1-arcsecond width when going through the sodium layer. Due to this width, and the thickness of the layer, the LGS has a cylinder shape on the sky, which generates a 3D focal volume after the telescope.

When using a Shack-Hartmann Wavefront Sensor (SHWFS), each subaperture observes the LGS from a different perspective. The subapertures closest to the laser launch telescope observe an object whose size is limited by the width of the laser beam, and the subapertures further away see an elongated shape, as they see this cylinder from the side. For a 40 m telescope, the geometry implies that the elongation on the furthest subaperture can reach up to 20 arcseconds. The measurements of the SHWFS rely on finding the displacement of the images of the source each subaperture generates. The extension of the source decreases the precision of the measurements, as it increases the observed variance of the position of the source. The extension of the source also means that large detectors have to be used to correctly sample the spot. An 80 x 80 subaperture SHWFS for a 40 m telescope would need a detector of around 1600 x 1600 pixels \citep{fusco2019story}. 

The pyramid wavefront sensor (PWFS; \citealp{1996JMOp...43..289R}) showed an interesting alternative to the SHWFS. It is from the family of the Fourier Filtering Wavefront Sensor (FFWFS), which uses the camera on the pupil plane. For this reason, an equivalent 80 x 80 subaperture PWFS would need a detector no bigger than 240 x 240 pixels. The PWFS also has a higher sensitivity than the SHWFS when using an NGS in closed loop operation, but as we showed in \citet{2024A&A...686A...1O}, the extension of the LGS, both due to the width of the laser beam, and the thickness of the sodium layer, lowered the sensitivity to a point where the performance of the AO loop was significantly degraded. This decrease in sensitivity comes from the fact that the extension of the source acts similarly to modulation.

In this paper, using the end-to-end physical optics models from OOMAO \citep{2014SPIE.9148E..6CC}, we compare the performance with respect to noise of the SHWFS and the PWFS to determine which is the best when using LGS. To do this, in Sec. \ref{sec:NoiseAndSens} we extended a sensitivity formalism from the PWFS to the SHWFS, to be able to compare the same metrics for both. Then, in Sec. \ref{sec:ClosedLoop} we show end-to-end simulations to validate the predictions of the noise model, and then use it to predict the noise propagation for 40 m telescopes. We then extend the analysis to compute the noise propagation for laser tomography adaptive optics (LTAO) systems with multiple WFSs.

\section{Noise and sensitivity}
\label{sec:NoiseAndSens}

\subsection{Pyramid wavefront sensor}

In this paper, we will use the reduced intensity approach presented in \citet{2023A&A...670A.153C} to compute the signal of the PWFS and the noise propagation. This reduced intensity is obtained by normalizing the intensity $I(\phi)$ recorded in the detector given an input phase $\phi$, by the number of photons in a frame $N_{ph}$, and then subtracting a reference intensity $I_0 = \frac{I(\phi = 0)}{N_{ph}}$ here chosen to be the intensity for a flat wavefront. With this, the reduced intensity is

\begin{equation}
    \Delta I(\phi) = \frac{I(\phi)}{N_{ph}} - I_0.
\end{equation}

To reconstruct the phase from a given WFS measurement we use an interaction matrix $\mathcal D = [\delta(\phi_1), \dots, \delta(\phi_N)]$, calibrated using a push-pull procedure inputting orthogonal modes (KL modes) in the phase space $[\phi_1, \dots, \phi_N]$ 

\begin{equation}
    \phi' = \mathcal D^\dagger \Delta I(\phi).
    \label{eq:phi_recon}
\end{equation}

As we record intensity information for given known phases, we have to invert the interaction matrix to recover the phase from the intensity measurements. For this, a pseudo-inverse approach is used, such that $\mathcal D^\dagger = (\mathcal D^t \mathcal D)^{-1} \, \mathcal D^t$.

PWFSs can operate with small, subelectron read-out noise (RON) detectors thanks to the limited number of pixel required. Nevertheless, to have the same conditions to compare with the SHWFS, we will use a RON value of $\Sigma_{RON} = 2 \, e^-/pix/frame$. To avoid confusion, in this work we will denote $\Sigma_{RON}$ as the standard deviation of the electronic noise of the detector, measured in $e^-/pix/frame$, and $\sigma^2_{RON}$ the phase variance introduced due to RON.  

Following \citet{2023A&A...670A.153C}, it is possible to define the sensitivity to RON $s(\phi_i)$, and to photon noise $s_\gamma(\phi_i)$, for each corrected mode $\phi_i$, such that the total phase variance introduced by noise $\sigma^2_{RON + \gamma}$ at each measurement is

\begin{equation}
\begin{split}
        \sigma^2_{RON} = \sum_N \frac{N_{sap} \, \Sigma_{RON}^2}{N_{ph}^2 \, s^2(\phi_i)}\\
        \sigma^2_\gamma = \sum_N \frac{1}{N_{ph} \, s_\gamma^2(\phi_i)}\\
        \sigma^2_{RON + \gamma} = \sigma^2_{RON} + \sigma^2_\gamma,
\end{split}
\label{eq:SigmaTotal}
\end{equation}

with $N_{sap}$ the number of subapertures. As a remark, the name sensitivity to RON (or photon noise) can be misleading, and sensitivity against RON (or photon noise) might be a better name. For consistency with the work developed by \citet{2023A&A...670A.153C}, we will use the former. For the control loop, we will use the same as in \citet{2024A&A...686A...1O}, such that the noise that is propagated through the loop is

\begin{equation}
    \hat \sigma^2_{RON + \gamma} = \delta \, \sigma^2_{RON + \gamma},
\end{equation}

with $\delta = 0.33$. In this work, we will use the \^. symbol to denote that the noise is propagated through the loop.

\subsection{Extending the sensitivity analysis to the Shack-Hartmann wavefront sensor}

There have been several works deriving theoretical formulas to predict the centroiding variance due to read-out and photon noise for a SHWFS for different centroiding algorithms \citep{1992A&A...261..677R,1999aoa..book...91R, 2006MNRAS.371..323T}, and some specifically for the prediction when using a laser guide star \citep{2010JOSAA..27A.201R}. In these papers, it is generally assumed that the source is diffraction-limited or it has a Gaussian profile. For both of these cases, analytical formulas have been derived considering the statistics of the noise.

These formulas are useful to observe the impact of the different design options of the AO system, but actual spot shapes might differ from the theoretical expectations. For example, in a real subaperture, a Gaussian object would not necessarily produce a Gaussian spot, as the image would also be convolved by the diffraction-limited spot. Elongated images of a laser guide star pose substantial difficulties, as each subaperture has a different image of the source. Also, LGS spots themselves might differ from Gaussian spots, given the density profile of the sodium layer.

With these issues in hand, we extended the sensitivity analysis developed for Fourier filtering WFS, presented in \citet{2023A&A...670A.153C}, to the SHWFS. The signal from the SHWFS will be processed using the Center of Gravity (CoG) or the Weighted CoG (WCoG, \citealp{2004OptL...29.2743N}) algorithm. We are interested in obtaining the same noise formalism as with the PWFS for the SHWFS, to be able to compare both using the same framework. Let us consider the following operation to reconstruct the phase using SHWFS measurements

\begin{equation}
    \phi' + \xi = \mathcal{D}_{SH}^\dagger \, CoG(I(\phi) + b(\phi)),
\end{equation}

where $\phi'$ is the phase estimation, $\xi$ the error in the phase reconstruction due to noise, $\mathcal{D}_{SH}$ is the interaction matrix, $CoG()$ is the center of gravity algorithm, $I(\phi)$ the noiseless intensity pattern in the detector and $b(\phi)$ the intensity due to RON and photon noise. If we assume that the proportion of light each subaperture receives is constant, it is possible to build a matrix $M$ that performs the center of gravity computation for all subapertures simultaneously if we normalize the intensity in the detector by the total number of photons in the frame

\begin{equation}
    CoG(I(\phi) + b(\phi)) = M \left(\frac{I(\phi) + b(\phi)}{N_{ph}}\right).
\end{equation}

Note that for WCoG we can construct the same matrix, but take into account the weighting mask, which can be different for each subaperture if needed. Also, when using WCoG, optical gains are needed to ensure a unitary response for all subapertures \citep{2006MNRAS.371..323T}. For this, we performed a calibration step where optical gains for the X and Y axis were computed for each subaperture, as this value changes depending on the elongation of the source.
With the $M$ matrix calibrated, we can express the phase reconstruction of the SHWFS as a series of matrix multiplications

\begin{equation}
    \phi + \xi =  \mathcal{D}_{SH}^\dagger M \frac{I(\phi) + b(\phi)}{N_{ph}}.
    \label{eq:SH_reconst}
\end{equation}

If we define $n(\phi) = b(\phi) / N_{ph}$ and $G^\dagger = \mathcal{D}_{SH}^\dagger\,  M$, the expression for the error in the phase reconstruction due to noise is 

\begin{equation}
    \xi = G^\dagger \, n(\phi).
\end{equation}

As we arrived at the same equation as Eq. 8 in \citet{2023A&A...670A.153C}, therefore we can follow the same steps to define both RON and photon noise sensitivities. For a given mode $\phi_i$, RON and photon noise sensitivities can be computed as
\begin{equation}
    \label{eq:sron_shwfs}
    s(\phi_i) = \sqrt{\frac{N_{sap}}{\left(G^\dagger G^{\dagger \, t}\right)_{i,i}}}
\end{equation}

\begin{equation}
    \label{eq:sgamma_sh}
    s_\gamma(\phi_i) = \frac{1}{\sqrt{\left(G^\dagger \, \textbf{diag}(I_0) \, G^{\dagger t}\right)_{i,i}}},
\end{equation}

with $N_{sap}$ the number of subapertures and $I_0 = \frac{I(\phi = 0)}{N_{ph}}$ the raw reference intensity of the detector of the SHWFS. Here we did not use the approximation $G^\dagger G^{\dagger \, t} = (G^t\,G)^\dagger$, therefore our expressions are slightly different but carry out the same meaning. With the sensitivities, we can then use Eq. \ref{eq:SigmaTotal} to compute the total residual phase variance $\sigma^2_{RON + \gamma}$ introduced by RON and photon noise at each measurement.

It is important to remark that $\sigma^2_{RON + \gamma}$ corresponds to the residual phase variance introduced at each measurement of the SHWFS, and not the centroiding error at each subaperture. Nevertheless, with this formalism, it is also possible to obtain the centroiding variance at each subaperture, by doing the same statistical analysis but without taking into account the interaction matrix, such that we obtain the raw centroids. Refer to Appendix \ref{App:CentrodingVariance} for the mathematical formalism and examples comparing the theoretical formulas with the predictions using this model. This could be useful, for example, if dealing with large simulations where the SHWFS model used is geometrical (i.e. computing directly the signal from the gradient of the incoming phase) instead of diffractive. In this case, the contributions of RON and photon noise to the centroiding variance can be precomputed by using a single noiseless image on the detector of the SHWFS.

\begin{figure}
    \centering
    \includegraphics[width = 0.49\textwidth]{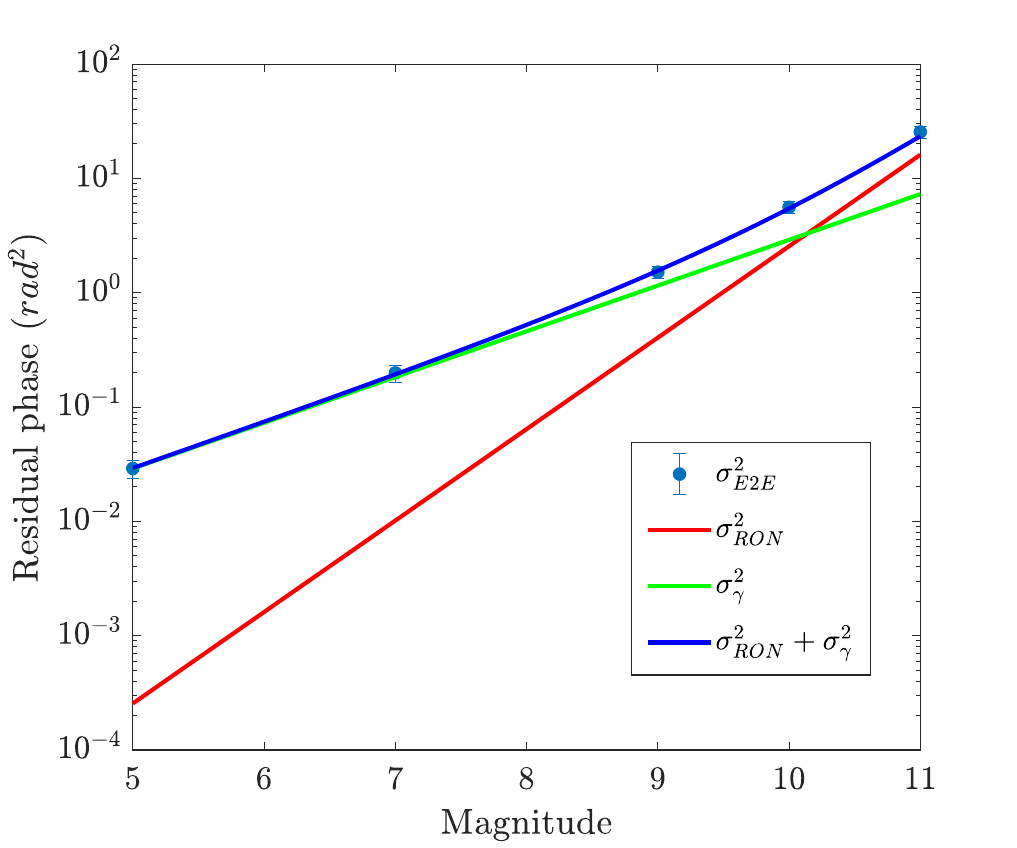}
    \caption{Evolution with respect to the magnitude of the guide star of the residual phase due to noise in open loop for a SHWFS with LGS for a 16 m telescope. The solid lines correspond to the residual phase variance due to RON (red line), photon noise (green line), and the sum of both (blue line), as predicted by the model. The markers correspond to the mean of 200 end-to-end iterations, with the errorbar the standard deviation of the residual variance. For reference, at 1 kHz and magnitude 10, $N_{ph} = 1.8 \times 10^5$ photons on the full pupil.}
    \label{fig:SH_OL_16m}
\end{figure}

To compare the sensitivity model with end-to-end simulations, we simulated 200 realizations with no atmosphere (i.e. a flat wavefront) of an open loop AO system with no controller for an SH with LGS for a 16 m telescope. We used the WCoG algorithm with a noiseless image of the detector as the weighting mask. For these simulations, we set the standard deviation of the read-out noise to be $2 \, e^-/pix/frame$. Figure \ref{fig:SH_OL_16m} shows with blue markers the result of the end-to-end simulations and with solid lines the expected residual variances predicted using the sensitivity model, where we found a good agreement between them. It is possible to observe that for low magnitudes the dominant term is photon noise (in green) and that at lower photon counts RON is the limiting factor (in red). We also repeated this for the regular CoG algorithm and found good agreement between end-to-end simulations and the predictions from the model.

\subsection{Comparing the sensitivities of both WFSs}

A previous work by \citet{2015OExpr..2328619P} explored the possibility of comparing both WFS with the same formalism, by using the Fisher Information matrix. In it, the authors develop the relationship between the Fisher coefficients and the noise propagation coefficients, making it possible to compare different wavefront sensors with the same metric. The main difference with this work is that for the SHWFS they use as signal the centroids, and for the centroiding variances they computed the theoretical values given the desired centroiding algorithm and the shape of the spot. In this work, we use as signal the raw image from the SHWFS and therefore we are not limited to the spot shapes the theoretical formulas have closed solutions for. With our formalism, the centroiding variance is a direct result (refer to App. \ref{appeq:cog_cov}) that is computed in parallel for all subapertures simultaneously. As we do not make any assumptions on the shape of the source, the sensitivity approach presented here is a more general formalism.

To compute these sensitivities we need access to the interaction matrix. For the PWFS, we used the same approach as in \citet{2024A&A...686A...1O}, where we computed the interaction matrix for the LGS by updating the optical gains of an interaction matrix built using a point source. By doing this the interaction matrix for the 40 m telescope with 81 x 81 actuators can be computed in a couple of hours.

\begin{figure}
    \centering
    \includegraphics[width = 0.49\textwidth]{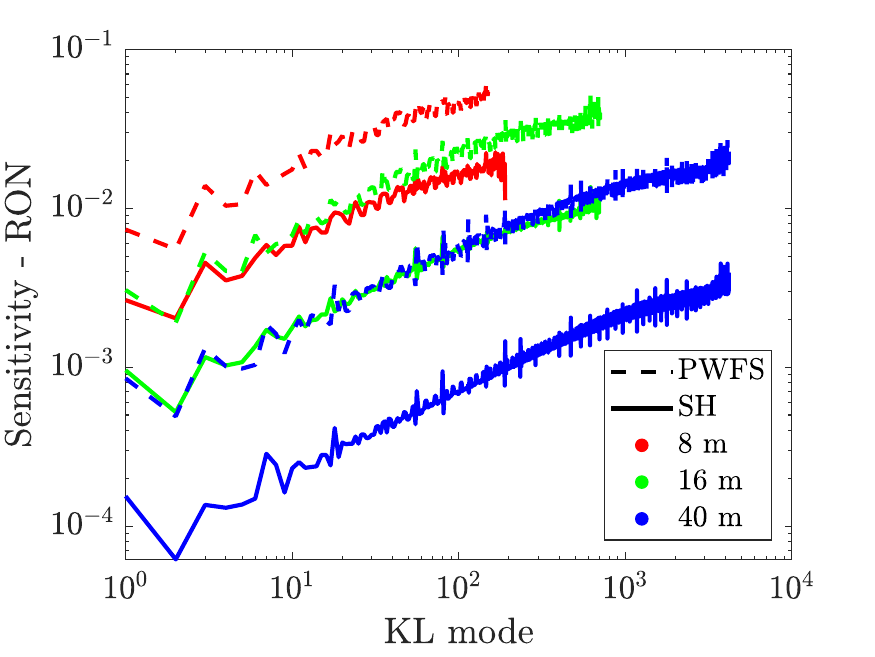}\\
    \includegraphics[width = 0.49\textwidth]{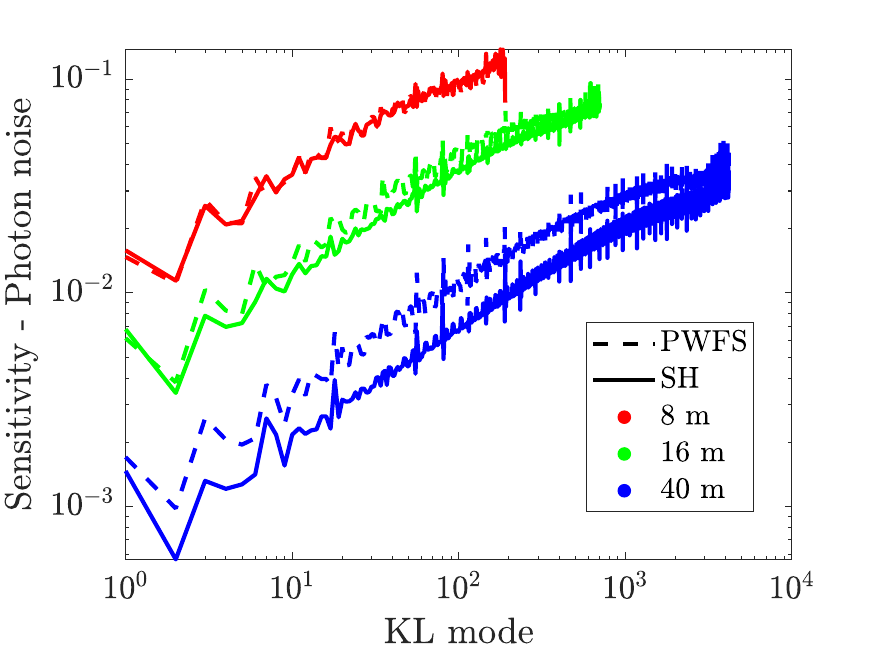}
    \caption{Sensitivity to RON (top plot) and photon noise (bottom plot) for an SHWFS (solid line) and a PWFS (dashed line) both using an LGS. The sensitivity was computed for telescopes of 8, 16 and 40 m in diameter (red, green and blue, respectively).}
    \label{fig:SH_sRON}
\end{figure}

The interaction matrix for the SHWFS for a 40 m telescope with 81 x 81 actuators can also take a long time to compute. For this reason, we took a shortcut on how to compute it. Let us assume that each actuator in the deformable mirror behaves identically, and the SHWFS has the same response to each actuator, just shifted in position. We can simulate a single push-pull operation on the central actuator, and then shift the signal of the WFS according the the position of each actuator. Once we have this, we will have access to the zonal interaction matrix. To get a modal interaction matrix, we can use a Zonal-to-Modal matrix, such that

\begin{equation}
    D_{modal} = D_{zonal} M2C
\end{equation}

the $M2C$ matrix can be obtained by computing the modes of the modal basis using the influence functions of the deformable mirror. Partially illuminated subapertures may not be perfectly modeled, but they still recover the wavefront accurately. With this approach, computing the modal interaction matrix takes minutes instead of weeks. Now that we have the modal interaction matrix, we can compute the sensitivities to RON and photon noise for WCoG, and compare them with the ones for the PWFS, as presented in Fig. \ref{fig:SH_sRON} where we computed the sensitivities for telescopes of 8, 16 and 40 m (red, green and blue curves, respectively). For both WFS we used the same number of subapertures, with a constant projected size of 50 cm. This meant that for the 8 m telescope, we used 16 subapertures, and for the 40 m, we used 80 subapertures. The pixel size on the SHWFS was 0.5 arcseconds (Nyquist sampling as the non-elongated axis has 1 arcsecond of FWHM). The weighting mask of each subaperture had a specific weighting function that corresponded to the convolution of a noiseless image from the detector with a Gaussian with 1 arcsecond of FWHM. This gives a good balance of retaining sensitivity and giving robustness to the system.

Regarding the sensitivity to RON (top figure), it is possible to observe that both the SHWFS (solid lines) and the PWFS (dashed lines) are more sensitive to higher-order modes than to lower-order ones. This is because both sensors are acting as gradient sensors: The SH by nature is a gradient sensor, and the PWFS, due to the size of the LGS is acting in the gradient-sensing regime \citep{2004OptCo.233...27V}. The first three modes used for this work are Tip, Tilt and focus, to then be able to filter them out in the closed loop to be corrected by an external loop. The rest of the modes correspond to KL modes built to be orthogonal to these three Zernike modes. When comparing the sensitivity to RON of the SH and the PWFS, the latter is consistently more sensitive, being on average 3 times more sensitive for the 8 m telescope and up to 5 times more for the 40 m one. The sensitivity itself is not the whole picture, as the detectors needed for the PWFS are small (only 240 x 240 pixels for an 80 x 80 subaperture PWFS), it is possible to use subelectron RON detectors \citep{2011aoel.confE..44G}, meaning that the value for $\Sigma^2_{RON}$ (and therefore $\sigma^2_{RON}$) in equation \ref{eq:SigmaTotal} would be much lower than for the SH.

Looking at the evolution of the sensitivity to RON with respect to the size of the telescope, it is possible to observe that both the SH and the PWFS have lower sensitivities as the diameter of the telescope increases. This can happen for several reasons: \textbf{(1)} for both WFS, to maintain the same equivalent diameter of the subapertures, bigger telescopes need more subapertures, which use more pixels. Increasing the number of pixels in the detector then decreases the sensitivity, as there is an increase in noisy measurements. \textbf{(2)} The elongation of the LGS increases with the diameter of the telescope. For the SH this means that the field of view of each subaperture has to increase, which increases the impact of noise on the measurement. For the PWFS, the elongation acts in a similar way to modulation, which reduces the sensitivity. 

The sensitivity to photon noise (bottom plot in Fig. \ref{eq:SigmaTotal}) is almost the same for both the SH and the PWFS for the 8 m telescope, and as the diameter increases the difference grows, as for the 16 m telescope the PWFS is around 1.2 times more sensitive and for the 40 m approximately 1.7 times more sensitive. For the 8 m telescope, the main structure of the laser beam that is limiting the sensitivity for both the PWFS and the SHWFS is the width of around 1 arcsecond, as the largest elongation is only around 2 to 3 arcseconds. As the telescope increases in diameter, the elongation of the spot of the laser is more pronounced in the SHWFS than in the PWFS given the depth of field: the subapertures of the SHWFS have the depth of field of a 50 cm diameter telescope, which is able to have almost all of the image of the LGS in focus. On the other hand, the PWFS has the depth of field of the full aperture, meaning that the image of the LGS quickly gets defocused, leaving the central focused spot as the limiting structure. For the 40 m telescope, this is around 2.5 arcseconds \citep{2024A&A...686A...1O}. 

The use of the WCoG algorithm is a trade-off between RON and photon noise, as increasing the size of the weighting mask reduces the propagation of photon noise, but increases that of the RON \citep{2004OptL...29.2743N}. If the RON value of a detector is known, it is possible to use this sensitivity analysis to optimize the size of the weighing mask. For example, if using a detector with low RON, it could be useful to use a weighing mask substantially larger than the noiseless spots. Not only it would help reduce the overall noise propagation, but also it could increase the linearity of the sensor. 

By observing both sensitivities, the advantage of the PWFS is mainly in the higher sensitivity to RON, as both have similar sensitivities to photon noise. The higher sensitivity to RON allows the PWFS to use more subapertures to measure the wavefront to mitigate aliasing, without sacrificing performance, as it will be limited by photon noise. 

\begin{figure}
    \centering
    \includegraphics[width = 0.49\textwidth]{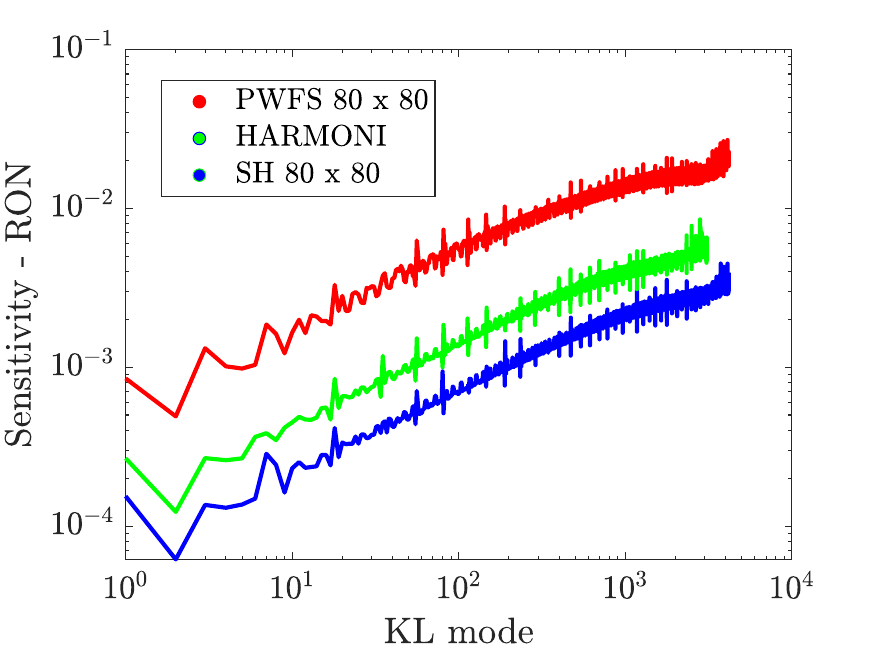}\\
    \includegraphics[width = 0.49\textwidth]{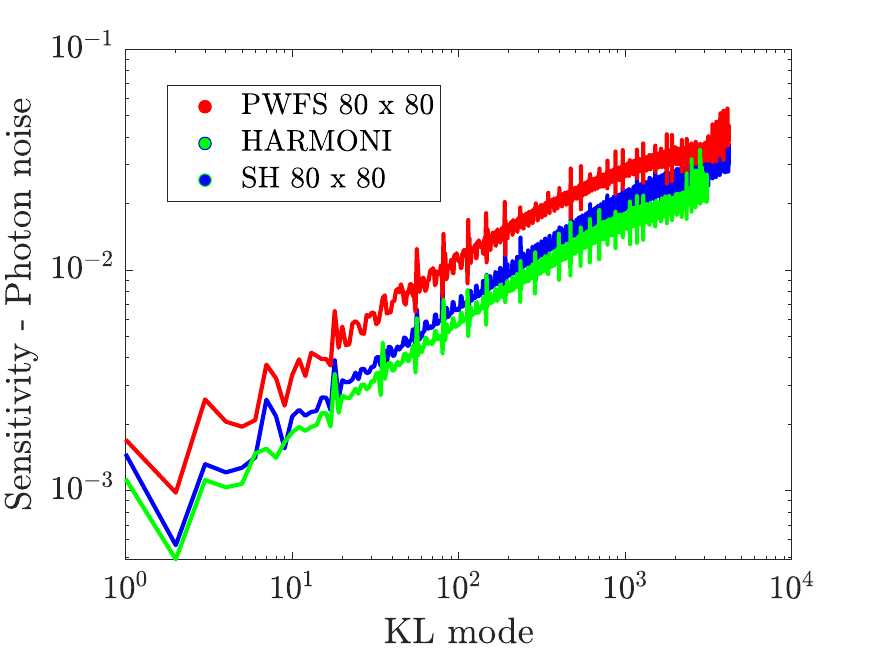}
    \caption{Comparison of the sensitivities of a PWFS (red) and a SHWFS (blue) with 80 x 80 subapertures, and a single SHWFS designed for HARMONI (green).}
    \label{fig:SH_HARMONI}
\end{figure}

In the case of the High Angular Resolution Monolithic Optical and Near-infrared Integral field spectrograph (HARMONI; \citealp{2016SPIE.9908E..1XT}), technological constraints forced the designers to lower the amount of subapertures from 80 x 80 to 68 x 68, and reduce the sampling from 0.5 $arcseconds/pix$ (i.e. Nyquist assuming a 1 arcsecond spot) to 1.15 $arcseconds/pix$. Cropping of the most elongated laser spots could also have an impact on the noise propagated, but it can not be represented by the model, therefore we will not take it into account. We used the sensitivity method to compare how noise behaves in a single SHWFS of HARMONI against an 80x80 subaperture SHWFS with 0.5 arcsecond pixels. Further analysis will be made in section \ref{sec:prediction40m} for the LTAO system, but as a first step we will observe the effects on a single WFS in a single conjugated adaptive optics (SCAO) configuration. Figure \ref{fig:SH_HARMONI} shows the sensitivities to RON and photon noise for a 40 m telescope using LGS with an 80 x 80 subaperture PWFS, an 80 x 80 subaperture SHWFS and a single 68 x 68 SHWFS designed for HARMONI in the SCAO configuration. As expected, the one designed for HARMONI has a higher sensitivity to RON, as it uses fewer subapertures and fewer pixels to make its measurements. In contrast, it has a lower sensitivity to photon noise. This could be due to the use of larger pixels, and therefore under-sampling, which leads to non-linearities and a loss in sensitivity \citep{2006MNRAS.371..323T}. 

\section{Closed Loop}
\label{sec:ClosedLoop}

\subsection{End-to-end simulations}

To observe the predictive capabilities of the sensitivity method, we did closed-loop simulations with 8 and 16 m telescopes for both a SHWFS and a PWFS and compared the results with those predicted by the model. The simulation parameters are found in table \ref{tab:params}. Here we chose a static atmosphere, such that the temporal errors due to wind did not affect the measurement of the contribution of RON and photon noise to the wavefront error. For the same reason, we chose the atmosphere to have a single ground layer, such that the cone effect would not affect the simulations, given that it is not correlated with noise. The cone effect will have a significant impact on the performance of the ELT, but to deal with this effect LTAO is planned to be used, but tomographic reconstruction of the wavefront is outside the scope of this work. We used the same sodium density profile for both WFS, that was presented in Fig. 2 of \citet{2014A&A...565A.102P}. We used a simple integrator in the feedback path with two frames of delay for the correction and a loop gain of 0.3. We also chose to include a simulation of a PWFS with an NGS, modulated at $4 \lambda/D$, to have a comparison for the LGS. The modulation radius was chosen to have a good trade-off between sensitivity and dynamic range. As the LGS can't measure global tip or tilt, and has difficulties measuring focus, a separate, noiseless AO loop was used to correct for these modes. For the science wavelength we chose the same as the sensing one for simplicity, as the difference between the performance curves, and therefore the comparison between the WFS, is independent on wavelength.

\begin{table}
    \caption{Simulation parameters}
    \label{tab:params}
    \centering
    \begin{tabular}{l l}
    \hline
    \hline
    \textbf{Telescopes}  &  \\
    Diameter     &        8.0, 16.0\\
    Throughput & 100 \%\\
    Central obstruction & None\\
    \\
    \textbf{Natural guide star}  &  \\
    Zenith angle & $0^o$\\
    Magnitudes & 5-16\\
    Zero point & $8.96 \times 10^9$ $photons/s/m^2$\\
    Modulation & $4 \, \lambda/D$\\
    \\
    
    \textbf{Laser guide star} & \\
    Launch geometry & Side launch \\
    Zenith angle & $0^o$\\
    Magnitudes & 5-14\\
    Zero point & $8.96 \times 10^9$ $photons/s/m^2$\\
    Number of samples & 10,000\\
    Sodium profile &  TopHatPeak \\
    \\
    \textbf{Atmosphere} & \\
    $r_0$ & 15 cm \\
    $L_0$ & 25 m\\ 
    Layers & 1\\
    Altitudes & 0 m\\
    Wind speed & 0 m/s\\
    \\
    \textbf{WFS} & \\
    Order for 8 m telescope & 16 x 16 subapertures\\
    Order for 16 m telescope & 32 x 32 subapertures\\
    Frequency & 1 KHz\\
    $\lambda_{sens}$ & 589 nm\\
    \\

    \textbf{SHWFS specific parameters} & \\
    Pixel scale & 0.5 arcseconds/pix \\
    Subaperture FoV for& \\
    8 m telescope  & 5 arcseconds\\
    & \\
    Subaperture FoV for & \\ 
    16 m telescope & 10 arcseconds\\
    \\
    
    \textbf{DM} & \\
    Order for 8 m telescope & 17 x 17 actuators \\
    & (200 KL modes) \\
    Order for 16 m telescope & 33 x 33 actuators \\
    & (700 KL modes) \\
    \\
    \textbf{AO loop} & \\
    Delay & 2 frames \\
    Gain & 0.3 \\
    \\
    \textbf{Science} & \\
    $\lambda_{sci}$ & 589 nm\\
    \hline
    \end{tabular}
\end{table}

A disadvantage that both WFS would have in common when using an LGS is the optical gain tracking (assuming a WCoG for the SHWFS), as the evolving structure of the sodium layer and changing seeing conditions will impact these values. Extra complexity both in software and hardware has to be included to compute the optical gains in real time. As LGS wavefront sensing is (mostly) blind to tip-tilt, a SHWFS system might introduce a tip-tilt mirror to modulate the LGS spots with a known displacement \citep{2004SPIE.5490.1384S}, such that it would be possible to estimate the optical gains by taking the ratio of the measured and known displacements. For the PWFS, the introduction of a gain scheduling camera as proposed in \citet{2021A&A...649A..70C} could allow for the online estimation of the optical gains. 

\begin{figure}
    \centering
    \includegraphics[width = 0.49\textwidth]{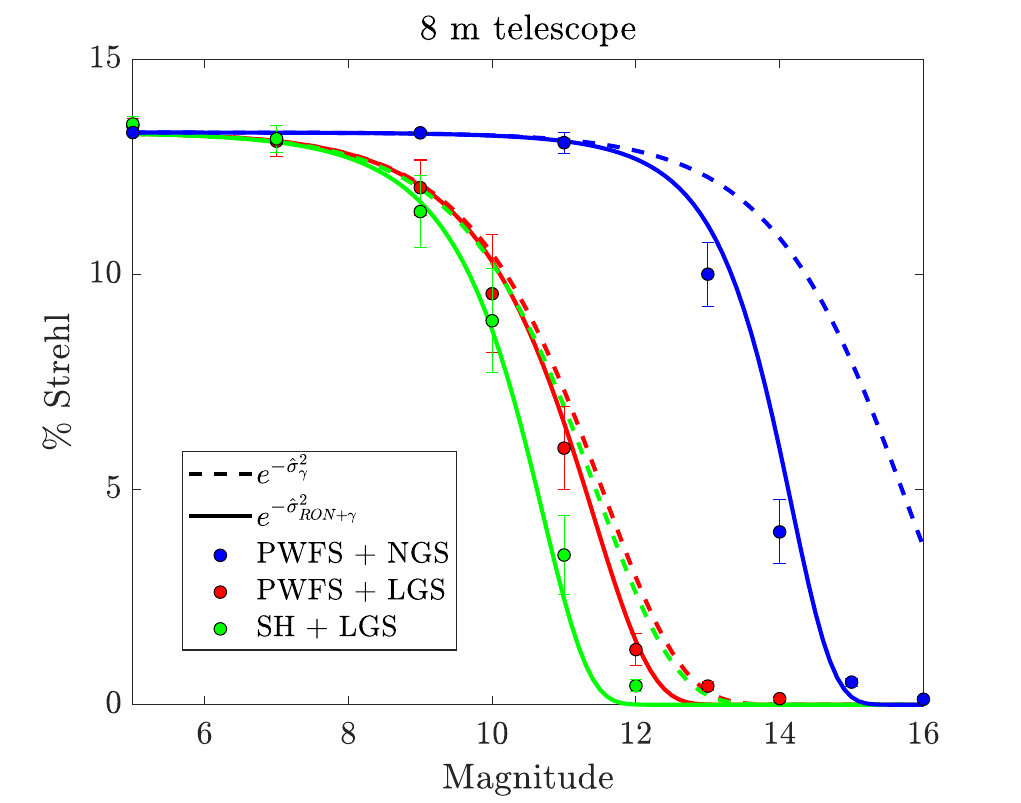}\\
    \includegraphics[width = 0.49\textwidth]{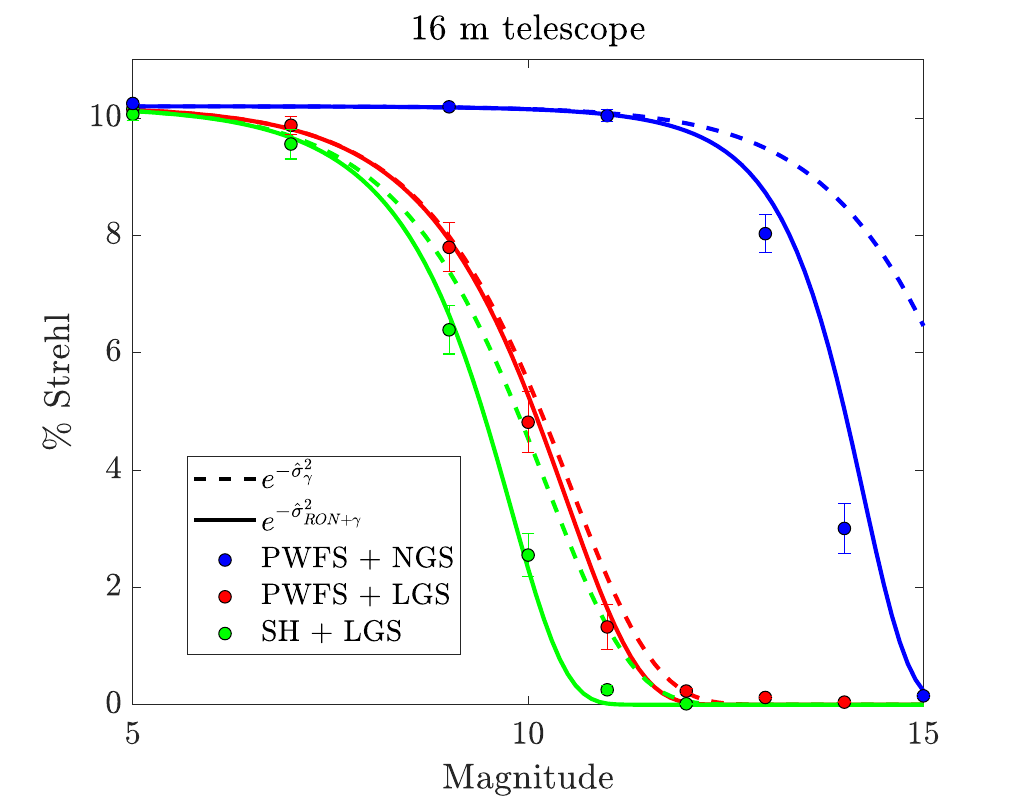}
    \caption{Performance of the AO loop for a SHWFS (green curves) and a PWFS (red curves) using LGS for an 8 (top) and 16 m telescopes (bottom). A PWFS using NGS (blue curve) was added as a reference for the performance.}
    \label{fig:CL}
\end{figure}

We tested several magnitudes for the guide stars and plotted the Strehl ratio obtained with the closed loop for each case in Fig. \ref{fig:CL}. For the residual phase due to fitting and aliasing, we simulated 20 phase realizations of a noiseless PWFS with NGS, and averaged the variance of the residual phase. The low performance is due to the low number of KL modes used to correct the wavefront. Fitting error alone limits the performance to around 20 \%, and aliasing or reconstruction error decreases the performance to the 10 - 15 \% range. 

Regarding aliasing, it has been shown that, when using an NGS, the PWFS has lower residual phase variance due to aliasing than the SH (assuming the same number of subapertures), given its flattening sensitivity curve \citep{2004OptCo.233...27V}. Nevertheless, for extended objects the size of the guide star has the same effect as modulation, which makes the PWFS operate as a gradient sensor, therefore we expect a similar behavior as for the SH. 

To obtain the curves of Fig. \ref{fig:CL} we first had to obtain the interaction matrices for each case. For the PWFS, we took the convolutional approach as in \citet{2024A&A...686A...1O}, where we computed the interaction matrix for an NGS, and then used a convolutional model \citep{2019JOSAA..36.1241F} to compute optical gains \citep{2008ApOpt..47...79K, 2018SPIE10703E..20D, 2021A&A...649A..70C} to optimize the interaction matrix for the LGS. With the interaction matrix, we could directly compute the sensitivities to RON and photon noise. For the SH, we computed the interaction matrix assuming that each actuator has the same signal footprint in the detector, just shifted depending on the position of the actuator. Then, we computed the optical gains to ensure a unitary response across all subapertures and finally build the matrix that performs the centroiding, which took into account the optical gains. With the interaction matrix and the centroiding matrix we could use equations \ref{eq:sron_shwfs} and \ref{eq:sgamma_sh} to compute the sensitivities to RON and photon noise, respectively.

Let us start analyzing only the theoretical performance expectations, represented as the curves in the figures, and not the E2E results, represented as the markers. Figure \ref{fig:CL} shows two curves for each tested case: the performance limited only by photon noise with a dashed curve, and the performance limited by both RON ($2 \, e^{-}/pix/frame$) and photon noise as a solid line. This distinction was made to understand which noise source is limiting the system. If the solid curve is close to the dashed, then the system would mainly be photon noise limited. On the other hand, if the solid curve deviates from the dashed, then the system would be limited mainly by RON. It is important to keep in mind that the evolution of noise propagated for RON and photon noise is different. For RON, residual phase variance goes as $N_{ph}^{-2}$ and for photon noise it goes as $N_{ph}^{-1}$. This means that, once the drop in performance starts (the "knee" in the plot), the fall is much steeper for RON than for photon noise. 

Being limited by one or the other noise has different implications. If the system is limited by photon noise, it means that it is possible to increase the number of pixels in the detector to improve aliasing or linearity, without degrading the noise performance, but improving the detector technology would not have a big impact on the noise propagated. On the other hand, if the system is limited by RON, then improving the technology (i.e. lowering the RON) would have a positive impact on the performance.

Specifically for the SH, if RON were dominating, then narrower weighting functions could be used at each subperture to increase the performance. On the other hand, if it was limited by photon noise (for example, using a detector with less RON), increasing the size of the weighting mask could improve the performance and linearity.

For the 8 m telescope, the PWFS with NGS, represented with the color blue, has the best noise performance. It can operate without a significant loss in performance up to magnitude 12. In this scenario, the solid line deviates from the dashed, meaning that RON is the limiting factor. In a real scenario, this would not be the case, because the low number of pixels allows for the use of subelectron RON detectors. Therefore, by improving the technology the performance can be similar to the dashed blue curve.

For the LGS, both the PWFS (in red) and the SH (in green) have a similar performance with respect to photon noise, as both dashed curves are nearly on top of each other. Nevertheless, even though both have the same RON, the higher sensitivity to RON of the PWFS makes it perform better than the SH. Using a detector with lower RON would not directly improve the noise performance of the PWFS with LGS, but it would allow to use more pixels (and therefore subapertures) to better sample the wavefront and help mitigate aliasing effects. The limiting source of noise for the SH is not as straightforward as with the PWFS. In this case, at high flux, the performance is limited by photon noise, as it is possible to observe that the beginning of the drop in performance follows the dashed curve. Then, as the photon flux diminishes, RON, with its steeper decrease dominates. This means that improving the technology of the detector would not improve the limiting magnitude, it would just improve the performance once it has already been affected by the return flux. 

For the 16 m telescope, the performance of the PWFS with NGS has almost the same evolution with magnitude as with the 8 m telescope. Having a bigger telescope gives a higher collection of photons, which reduces the overall noise propagated. Nevertheless, the bigger telescope needs more subapertures to have the same pitch as the smaller one. These two effects cancel out each other, making the overall performance approximately independent on the telescope size. 

For the LGS, the expected performance of both the PWFS and the SH are similar to that of the 8 m telescope, but the curves are shifted to the left around 1 magnitude. Given the same arguments as for the NGS, bigger telescopes gather more light, but need more pixels and subapertures to maintain the same pitch. On top of that, the elongation of the LGS is increased by the size of the telescope: the bigger the elongation the lower the sensitivity, therefore a worse performance. Looking only at the PWFS, it is possible to observe that the solid curve is almost the same as the dashed, therefore photon noise is the limiting factor. This means that increasing the number of pixels and therefore subapertures is a possible options to help mitigate aliasing. For the SH, again the limiting magnitude is given by photon noise, as the "knee" in the plot is the same as the dashed curve, and then at lower photon fluxes RON starts to dominate. 

Now observing the markers, for the PWFS with NGS it is possible to observe a good agreement between the theoretical expectations and the E2E simulations. The drop in performance is around 0.3 magnitudes less for the E2E simulations than what was expected with the model. This can be explained as the sensitivity model does not take into account the non-linearities of the PWFS, which are known to decrease the sensitivity \citep{2020A&A...644A...6C}. For the extended objects, both the PWFS and SH show good agreement between the E2E simulations and the theoretical curves.

\subsection{Prediction of the performance of the 40 m telescope}
\label{sec:prediction40m}

\begin{figure}
    \centering
    \includegraphics[width = 0.49\textwidth]{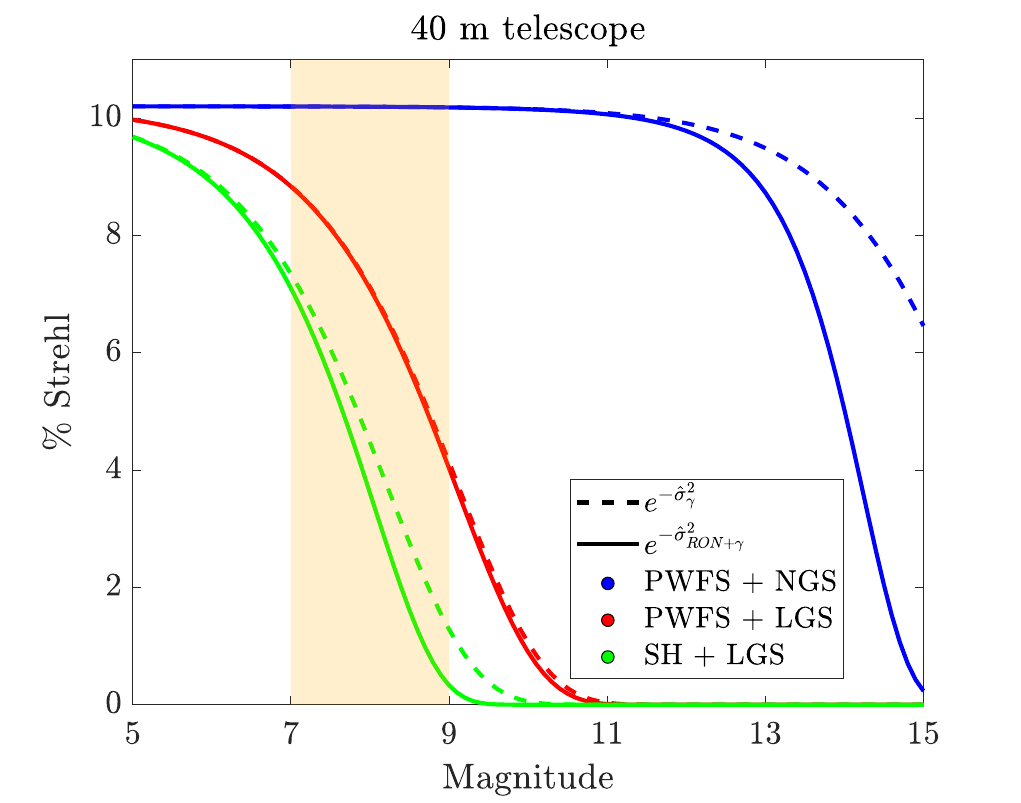}
    \caption{Expected performance of the AO loop for a SHWFS and a PWFS using LGS for a 40 m telescope. A PWFS using NGS was added as a reference for the performance.}
    \label{fig:CL_40m}
\end{figure}

Having shown that the sensitivity model accurately predicts the performance of the AO loop, we can now use it for the 40 m telescope for both WFS. These correspond to 80 x 80 subapertures, with an 81 x 81 deformable mirror with 5100 actuators. The sampling of the SHWFS is 0.5 $arcseconds/pix$ with a field of view of 25 arcseconds. A detector of this size and noise characteristics is not available, but as we showed in figure \ref{fig:SH_HARMONI} we expect a similar performance as if using the detector designed for HARMONI. The weighting mask used was a noiseless image of the detector, convolved with a 0.5 arcseconds FWHM Gaussian for purposes of robustness and linearity. Figure \ref{fig:CL_40m} shows the expected performance for both the SHWFS and the PWFS for a 40 m telescope. A yellow vertical stripe was included in the plot as a reference for the expected return of the LGS. 

Considering the chosen design parameters, the PWFS is photon noise limited, and the SH is mainly photon noise limited, but RON has an impact at lower fluxes. The PWFS has a limiting magnitude around one magnitude higher than the SH, and the fact that it is photon noise limited means that increasing the number of pixels could increase the performance with respect to aliasing, without sacrificing performance with respect to noise. Nevertheless, at the expected return fluxes for the LGS (magnitudes between 7 and 9), both the PWFS and the SH are not only in a low-performance region but also small changes in the return flux would result in large variations in performance. 

Considering the SH designed for HARMONI, as it has a higher sensitivity to RON and a slightly lower sensitivity to photon noise than the SH tested here (refer to Fig. \ref{fig:SH_HARMONI}), we expect it to behave in a similar way to the well sampled, 80 x 80 subaperture SH.

\subsection{Expected noise performance in a tomographic configuration for a 40 m telescope}

In general, the use of LGS for wavefront sensing is not done with a single guide source, but several of these are implemented to, for example, use wide-field adaptive optics, or mitigate cone effect. For this, Laser Tomography Adaptive Optics (LTAO) or Multi Conjugated Adaptive Optics (MCAO) systems are used. 

In the following discussion, we will estimate the noise performance of an LTAO system. Tomographic reconstructors pose substantial difficulties in estimating the propagation of noise, therefore we will use only a simplified version. We will assume each WFS observes the same atmosphere and has uncorrelated RON and photon noise. This approach is not useful for the tomographic reconstruction of the atmosphere, but it can give insights into the expected performance with respect to noise. Using LTAO also allows for the use of super-resolution \citep{2022A&A...667A..48O}, which could help mitigate aliasing. This will not be taken into account for the discussion, therefore the value of the Strehl ratio at high fluxes is not representative of a full LTAO reconstruction, but the evolution with respect to noise is.

Let us assume an LTAO system with 6 LGS and 6 WFS. For both wavefront sensors, each subaperture sees a different perspective of the LGS, elongated and oriented given the position of the subaperture in the pupil and the location of the laser launch telescope. This implies that each subaperture will produce a measurement (centroiding for the SH and slopes for the PWFS)  with a variance that is dependent on its position in the pupil. Similar to what has been done in \citet{2010SPIE.7736E..0XT} and \citet{2012SPIE.8447E..2CB}, in Appendix \ref{App:multipleLGS} we show that by using a Generalized least square estimator (GLSE) to combine the measurements at each subaperture produced by the LGSs, the overall variance of the signal at each subaperture is better than using a single WFS with a non-elongated 1-arcsecond spot. Table \ref{tab:SignalVariances} shows the average signal variances for both wavefront sensors using 6 LGS, combining the signal using the GLSE. The variances are normalized by the signal variance of a non-elongated 1-arcsecond spot. We used the results from the GLSE for the noise propagation as an approximation of the multiple LGS system.

\begin{table*}
\caption{Average variance of the signal coming from the SHWFS and PWFS for a 40 m telescope with 6 LGS using the generalized least squares estimator.}
    \centering
    \begin{tabular}{l l l}
    \hline \hline 
    \textbf{Wavefront sensor}    &  \textbf{Normalized average signal} & \textbf{Normalized average signal}\\
     & \textbf{variance due to RON } & \textbf{variance due to Photon noise}\\ 
     6 SH + 6 LGS & 0.81 & 0.24\\
     6 PWFS + 6 LGS & 0.70 & 0.50\\\hline

    \end{tabular}
    \label{tab:SignalVariances}
    \tablefoot{The variances are normalized by the variance obtained by using a 1-arcsecond spot as a guide source.}
\end{table*}

\begin{figure}
    \centering
    \includegraphics[width = 0.49\textwidth]{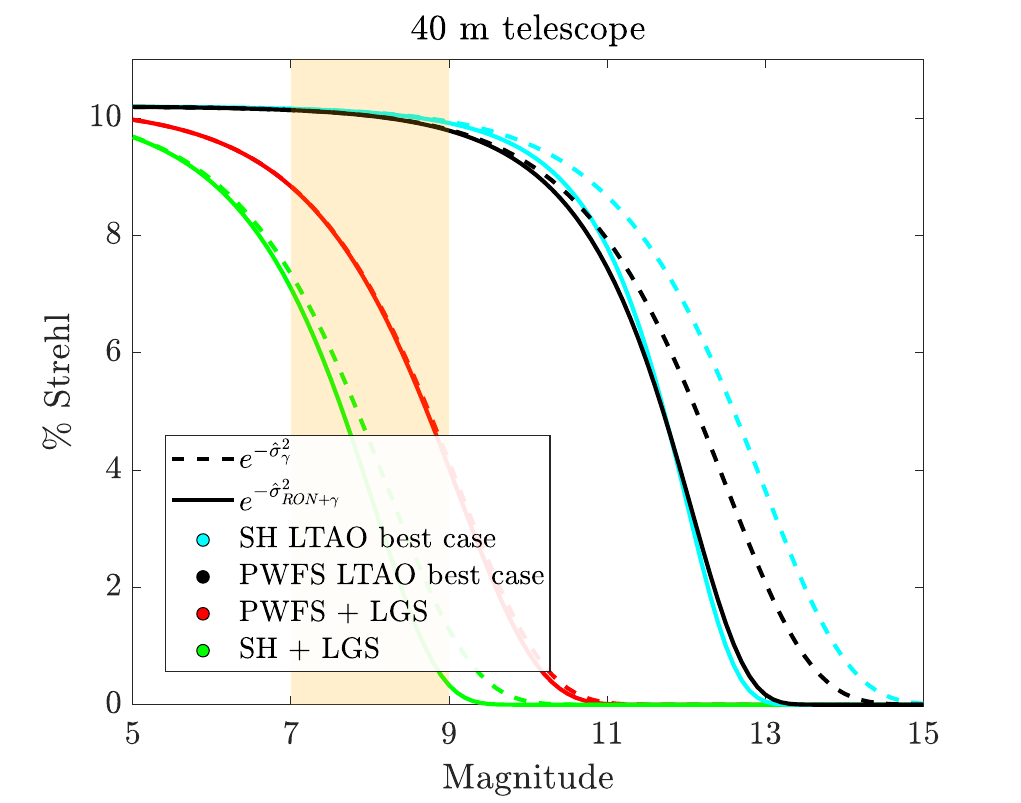}
    \caption{Performance of the AO loop for the best case LTAO using PWFS (in black) and SHWFS (in cyan). In red and green is the performance of a single PWFS and SHWFS, respectively.}
    \label{fig:CL_LTAO}
\end{figure}

Figure \ref{fig:CL_LTAO} shows the expected performance of these approximated systems. The increase in performance of the SHWFS is drastic, as the limiting magnitude of the best-case scenario (cyan curve) is 4.4 magnitudes greater than for the single WFS case. For the PWFS, the increase in performance is of 3.2 magnitudes (black curve). By combining the measurements, the SHWFS could be capable of outperforming the PWFS by 0.3 magnitudes, requiring about 25 \% less photons to achieve the same performance. Nevertheless, both WFS would be able to operate without any significant loss in performance at the expected return fluxes of the LGS. As the SH is affected more by RON than the PWFS, we increased the value and found that the PWFS system (in a multiple WFS configuration) can perform better than the SH system if RON is higher than $7 \, e^-/pix/frame$. But, as the pixel requirements of the PWFS are low, it can use a sub-electron RON detector, and in that case the performances would be almost identical (dashed black curve in figure \ref{fig:CL_LTAO}).


A Pyramid-like WFS could be used with the geometry of the LGS into consideration, as in for example the Ingot WFS proposed by \citet{2024A&A...688A..21R}. As it follows the height extension of the LGS, it could operate as if it were a regular PWFS with a 1-arcsecond spot. In this ideal scenario, the equivalent 6 Ingot + 6 LGS with the generalized least square estimator could outperform the SH system by around 1 magnitude.


\section{Conclusion}

In this work, we computed the expected performance of the AO loop for the SHWFS and the PWFS, by using an analytical model.

To do this, we extended a sensitivity analysis developed for the PWFS to the SHWFS by expressing the centroiding algorithm as a matrix multiplication. This allowed us to use the same formalism for both wavefront sensors and use the same metrics for a fair comparison between the two.

We found that the main difference between the two wavefront sensors is in the sensitivity to RON, in which the PWFS has an advantage. Sensitivity to photon noise is the same for both in an 8 m telescope and 1.7 times higher for the PWFS at 40 m. 

As photon noise is what is limiting the performance of the PWFS, it could be possible to increase the number of subapertures to mitigate aliasing. For the SHWFS both RON and photon noise are limiting its performance, therefore it is not advisable to increase the number of subapertures.

Approximating the noise propagation of the LTAO system by using a generalized least squares estimator, we found that the multiple SH system has a limiting magnitude up to 4.4 magnitudes higher than the single WFS system. For the PWFS, it would increase the limiting magnitude by 3.2. In this scenario, the SH system has a limiting magnitude 0.3 higher than the PWFS system. If using a sub-electron RON detector for the PWFS, both systems would perform almost identically. Nevertheless, if a WFS designed specifically for the LGS geometry is capable of achieving the performance of a PWFS with a non-elongated 1-arcsecond spot, it could outperform the SH system by 1 magnitude.

If the LTAO reconstruction is capable of performing as the approximated model either WFS would have almost the same performance with respect to noise.


\begin{acknowledgements}
This work benefited from the support of the the French National Research Agency (ANR) with \emph{WOLF (ANR-18-CE31-0018)}, \emph{APPLY (ANR-19-CE31-0011)} and \emph{LabEx FOCUS (ANR-11-LABX-0013)}; the Programme Investissement Avenir \emph{F-CELT (ANR-21-ESRE-0008)}, the \emph{Action Spécifique Haute Résolution Angulaire (ASHRA)} of CNRS/INSU co-funded by CNES, the \emph{ECOS-CONYCIT} France-Chile cooperation (\emph{C20E02}), the \emph{ORP-H2020} Framework Programme of the European Commission’s (Grant number \emph{101004719}), \emph{STIC AmSud (21-STIC-09)}, the french government under the \emph{France 2030 investment plan}, as part of the \emph{Initiative d'Excellence d'Aix-Marseille Université A*MIDEX, program number AMX-22-RE-AB-151}, the \emph{Conseil régional Provence-Alpes-Côte d'Azur} with the \emph{emplois jeune doctorant} program, co-funded by \emph{First Light Imaging}, and the Millennium Science Initiative Program (ACIP, NCN19 161).
\end{acknowledgements}

\bibliographystyle{aa}
\bibliography{bibliography}

\begin{thebibliography}{30}
\expandafter\ifx\csname natexlab\endcsname\relax\def\natexlab#1{#1}\fi

\bibitem[{{B{\'e}chet} {et~al.}(2012){B{\'e}chet}, {Tallon}, \& {Thi{\'e}baut}}]{2012SPIE.8447E..2CB}
{B{\'e}chet}, C., {Tallon}, M., \& {Thi{\'e}baut}, {\'E}. 2012, in Society of Photo-Optical Instrumentation Engineers (SPIE) Conference Series, Vol. 8447, Adaptive Optics Systems III, ed. B.~L. {Ellerbroek}, E.~{Marchetti}, \& J.-P. {V{\'e}ran}, 84472C

\bibitem[{{Chambouleyron} {et~al.}(2020){Chambouleyron}, {Fauvarque}, {Janin-Potiron}, {Correia}, {Sauvage}, {Schwartz}, {Neichel}, \& {Fusco}}]{2020A&A...644A...6C}
{Chambouleyron}, V., {Fauvarque}, O., {Janin-Potiron}, P., {et~al.} 2020, \aap, 644, A6

\bibitem[{{Chambouleyron} {et~al.}(2023){Chambouleyron}, {Fauvarque}, {Plantet}, {Sauvage}, {Levraud}, {Ciss{\'e}}, {Neichel}, \& {Fusco}}]{2023A&A...670A.153C}
{Chambouleyron}, V., {Fauvarque}, O., {Plantet}, C., {et~al.} 2023, \aap, 670, A153

\bibitem[{{Chambouleyron} {et~al.}(2021){Chambouleyron}, {Fauvarque}, {Sauvage}, {Neichel}, \& {Fusco}}]{2021A&A...649A..70C}
{Chambouleyron}, V., {Fauvarque}, O., {Sauvage}, J.~F., {Neichel}, B., \& {Fusco}, T. 2021, \aap, 649, A70

\bibitem[{{Ciliegi} {et~al.}(2022){Ciliegi}, {Agapito}, {Aliverti}, {Annibali}, {Arcidiacono}, {Azzaroli}, {Balestra}, {Baronchelli}, {Baruffolo}, {Bergomi}, {Bianco}, {Bonaglia}, {Briguglio}, {Busoni}, {Cantiello}, {Capasso}, {Carl{\`a}}, {Carolo}, {Cascone}, {Chinellato}, {Cianniello}, {Colapietro}, {Correia}, {Cosentino}, {D'Auria}, {De Caprio}, {Devaney}, {Di Antonio}, {Di Cianno}, {Di Dato}, {Di Giammatteo}, {Di Rico}, {Dolci}, {Eredia}, {Esposito}, {Fantinel}, {Farinato}, {Feautrier}, {Foppiani}, {Genoni}, {Giro}, {Gluck}, {Goncharov}, {Grani}, {Greggio}, {Guieu}, {Gullieuszik}, {Haguenauer}, {Hubert}, {Lapucci}, {Laudisio}, {Le Louarn}, {Magrin}, {Malone}, {Marafatto}, {Munari}, {Oberti}, {Pariani}, {Pettazzi}, {Plantet}, {Portaluri}, {Puglisi}, {Rabou}, {Ragazzoni}, {Redaelli}, {Riva}, {Rochat}, {Rodeghiero}, {Salasnich}, {Savarese}, {Scalera}, {Schipani}, {Sordo}, {Sztefek}, {Valentini}, \& {Xompero}}]{2022SPIE12185E..14C}
{Ciliegi}, P., {Agapito}, G., {Aliverti}, M., {et~al.} 2022, in Society of Photo-Optical Instrumentation Engineers (SPIE) Conference Series, Vol. 12185, Adaptive Optics Systems VIII, ed. L.~{Schreiber}, D.~{Schmidt}, \& E.~{Vernet}, 1218514

\bibitem[{{Conan} \& {Correia}(2014)}]{2014SPIE.9148E..6CC}
{Conan}, R. \& {Correia}, C. 2014, in Society of Photo-Optical Instrumentation Engineers (SPIE) Conference Series, Vol. 9148, Adaptive Optics Systems IV, ed. E.~{Marchetti}, L.~M. {Close}, \& J.-P. {Vran}, 91486C

\bibitem[{{Davies} {et~al.}(2016){Davies}, {Schubert}, {Hartl}, {Alves}, {Cl{\'e}net}, {Lang-Bardl}, {Nicklas}, {Pott}, {Ragazzoni}, {Tolstoy}, {Agocs}, {Anwand-Heerwart}, {Barboza}, {Baudoz}, {Bender}, {Bizenberger}, {Boccaletti}, {Boland}, {Bonifacio}, {Briegel}, {Buey}, {Chapron}, {Cohen}, {Czoske}, {Dreizler}, {Falomo}, {Feautrier}, {F{\"o}rster Schreiber}, {Gendron}, {Genzel}, {Gl{\"u}ck}, {Gratadour}, {Greimel}, {Grupp}, {H{\"a}user}, {Haug}, {Hennawi}, {Hess}, {H{\"o}rmann}, {Hofferbert}, {Hopp}, {Hubert}, {Ives}, {Kausch}, {Kerber}, {Kravcar}, {Kuijken}, {Lang-Bardl}, {Leitzinger}, {Leschinski}, {Massari}, {Mei}, {Merlin}, {Mohr}, {Monna}, {M{\"u}ller}, {Navarro}, {Plattner}, {Przybilla}, {Ramlau}, {Ramsay}, {Ratzka}, {Rhode}, {Richter}, {Rix}, {Rodeghiero}, {Rohloff}, {Rousset}, {Ruddenklau}, {Schaffenroth}, {Schlichter}, {Sevin}, {Stuik}, {Sturm}, {Thomas}, {Tromp}, {Turatto}, {Verdoes-Kleijn}, {Vidal}, {Wagner}, {Wegner}, {Zeilinger}, {Ziegler}, \& {Zins}}]{2016SPIE.9908E..1ZD}
{Davies}, R., {Schubert}, J., {Hartl}, M., {et~al.} 2016, in Society of Photo-Optical Instrumentation Engineers (SPIE) Conference Series, Vol. 9908, Ground-based and Airborne Instrumentation for Astronomy VI, ed. C.~J. {Evans}, L.~{Simard}, \& H.~{Takami}, 99081Z

\bibitem[{{Deo} {et~al.}(2018){Deo}, {Gendron}, {Rousset}, {Vidal}, \& {Buey}}]{2018SPIE10703E..20D}
{Deo}, V., {Gendron}, {\'E}., {Rousset}, G., {Vidal}, F., \& {Buey}, T. 2018, in Society of Photo-Optical Instrumentation Engineers (SPIE) Conference Series, Vol. 10703, Adaptive Optics Systems VI, ed. L.~M. {Close}, L.~{Schreiber}, \& D.~{Schmidt}, 1070320

\bibitem[{{Fauvarque} {et~al.}(2019){Fauvarque}, {Janin-Potiron}, {Correia}, {Br{\^u}l{\'e}}, {Neichel}, {Chambouleyron}, {Sauvage}, \& {Fusco}}]{2019JOSAA..36.1241F}
{Fauvarque}, O., {Janin-Potiron}, P., {Correia}, C., {et~al.} 2019, Journal of the Optical Society of America A, 36, 1241

\bibitem[{{Foy} \& {Labeyrie}(1985)}]{1985A&A...152L..29F}
{Foy}, R. \& {Labeyrie}, A. 1985, \aap, 152, L29

\bibitem[{{Fusco} {et~al.}(2019){Fusco}, {Neichel}, {Correia}, {Blanco}, {Costille}, {Dohlen}, {Rigaut}, {Renaud}, {Bonnefoi}, Ke, {et~al.}}]{fusco2019story}
{Fusco}, T., {Neichel}, B., {Correia}, C., {et~al.} 2019, in AO4ELT6

\bibitem[{{Gach} {et~al.}(2011){Gach}, {Balard}, {Stadler}, {Guillaume}, \& {Feautrier}}]{2011aoel.confE..44G}
{Gach}, J.-L., {Balard}, P., {Stadler}, E., {Guillaume}, C., \& {Feautrier}, P. 2011, in Second International Conference on Adaptive Optics for Extremely Large Telescopes. Online at <A href=``http://ao4elt2.lesia.obspm.fr''>http://ao4elt2.lesia.obspm.fr</A, 44

\bibitem[{{Gilmozzi} \& {Spyromilio}(2007)}]{2007Msngr.127...11G}
{Gilmozzi}, R. \& {Spyromilio}, J. 2007, The Messenger, 127, 11

\bibitem[{{Korkiakoski} {et~al.}(2008){Korkiakoski}, {V{\'e}rinaud}, \& {Louarn}}]{2008ApOpt..47...79K}
{Korkiakoski}, V., {V{\'e}rinaud}, C., \& {Louarn}, M.~L. 2008, \ao, 47, 79

\bibitem[{{Nicolle} {et~al.}(2004){Nicolle}, {Fusco}, {Rousset}, \& {Michau}}]{2004OptL...29.2743N}
{Nicolle}, M., {Fusco}, T., {Rousset}, G., \& {Michau}, V. 2004, Optics Letters, 29, 2743

\bibitem[{{Oberti} {et~al.}(2022){Oberti}, {Correia}, {Fusco}, {Neichel}, \& {Guiraud}}]{2022A&A...667A..48O}
{Oberti}, S., {Correia}, C., {Fusco}, T., {Neichel}, B., \& {Guiraud}, P. 2022, \aap, 667, A48

\bibitem[{{Oyarz{\'u}n} {et~al.}(2024){Oyarz{\'u}n}, {Chambouleyron}, {Neichel}, {Fusco}, \& {Guesalaga}}]{2024A&A...686A...1O}
{Oyarz{\'u}n}, F., {Chambouleyron}, V., {Neichel}, B., {Fusco}, T., \& {Guesalaga}, A. 2024, \aap, 686, A1

\bibitem[{{Pfrommer} \& {Hickson}(2014)}]{2014A&A...565A.102P}
{Pfrommer}, T. \& {Hickson}, P. 2014, \aap, 565, A102

\bibitem[{{Plantet} {et~al.}(2015){Plantet}, {Meimon}, {Conan}, \& {Fusco}}]{2015OExpr..2328619P}
{Plantet}, C., {Meimon}, S., {Conan}, J.~M., \& {Fusco}, T. 2015, Optics Express, 23, 28619

\bibitem[{{Primmerman} {et~al.}(1991){Primmerman}, {Murphy}, {Page}, {Zollars}, \& {Barclay}}]{1991Natur.353..141P}
{Primmerman}, C., {Murphy}, D., {Page}, D., {Zollars}, B., \& {Barclay}, H. 1991, \nat, 353, 141

\bibitem[{{Ragazzoni}(1996)}]{1996JMOp...43..289R}
{Ragazzoni}, R. 1996, Journal of Modern Optics, 43, 289

\bibitem[{{Ragazzoni} {et~al.}(2024){Ragazzoni}, {Portaluri}, {Greggio}, {Dima}, {Arcidiacono}, {Bergomi}, {Di Filippo}, {Gomes Machado}, {Santhakumari}, {Viotto}, {Battaini}, {Carolo}, {Chinellato}, {Farinato}, {Magrin}, {Marafatto}, {Umbriaco}, \& {Vassallo}}]{2024A&A...688A..21R}
{Ragazzoni}, R., {Portaluri}, E., {Greggio}, D., {et~al.} 2024, \aap, 688, A21

\bibitem[{{Rigaut} \& {Gendron}(1992)}]{1992A&A...261..677R}
{Rigaut}, F. \& {Gendron}, E. 1992, \aap, 261, 677

\bibitem[{{Robert} {et~al.}(2010){Robert}, {Conan}, {Gratadour}, {Schreiber}, \& {Fusco}}]{2010JOSAA..27A.201R}
{Robert}, C., {Conan}, J.-M., {Gratadour}, D., {Schreiber}, L., \& {Fusco}, T. 2010, Journal of the Optical Society of America A, 27, A201

\bibitem[{{Rousset}(1999)}]{1999aoa..book...91R}
{Rousset}, G. 1999, in Adaptive Optics in Astronomy, ed. F.~{Roddier}, 91

\bibitem[{{Saddlemyer} {et~al.}(2004){Saddlemyer}, {Herriot}, {Vrran}, {Smith}, \& {Dunn}}]{2004SPIE.5490.1384S}
{Saddlemyer}, L.~K., {Herriot}, G., {Vrran}, J.-P., {Smith}, M., \& {Dunn}, J. 2004, in Society of Photo-Optical Instrumentation Engineers (SPIE) Conference Series, Vol. 5490, Advancements in Adaptive Optics, ed. D.~{Bonaccini Calia}, B.~L. {Ellerbroek}, \& R.~{Ragazzoni}, 1384--1392

\bibitem[{{Tallon} {et~al.}(2010){Tallon}, {Tallon-Bosc}, {B{\'e}chet}, {Momey}, {Fradin}, \& {Thi{\'e}baut}}]{2010SPIE.7736E..0XT}
{Tallon}, M., {Tallon-Bosc}, I., {B{\'e}chet}, C., {et~al.} 2010, in Society of Photo-Optical Instrumentation Engineers (SPIE) Conference Series, Vol. 7736, Adaptive Optics Systems II, ed. B.~L. {Ellerbroek}, M.~{Hart}, N.~{Hubin}, \& P.~L. {Wizinowich}, 77360X

\bibitem[{{Thatte} {et~al.}(2016){Thatte}, {Clarke}, {Bryson}, {Shnetler}, {Tecza}, {Fusco}, {Bacon}, {Richard}, {Mediavilla}, {Neichel}, {Arribas}, {Garcia-Lorenzo}, {Evans}, {Remillieux}, {El Madi}, {Herreros}, {Melotte}, {O'Brien}, {Tosh}, {Vernet}, {Hammersley}, {Ives}, {Finger}, {Houghton}, {Rigopoulou}, {Lynn}, {Allen}, {Zieleniewski}, {Kendrew}, {Ferraro-Wood}, {P{\'e}contal-Rousset}, {Kosmalski}, {Laurent}, {Loupias}, {Piqueras}, {Renault}, {Blaizot}, {Daguis{\'e}}, {Migniau}, {Jarno}, {Born}, {Gallie}, {Montgomery}, {Henry}, {Schwartz}, {Taylor}, {Zins}, {Rodr{\'\i}guez-Ramos}, {Cagigas}, {Battaglia}, {Rebolo L{\'o}pez}, {Hern{\'a}ndez Su{\'a}rez}, {Gigante-Ripoll}, {Piqueras L{\'o}pez}, {Villar Martin}, {Correia}, {Pascal}, {Blanco}, {Vola}, {Epinat}, {Peroux}, {Vigan}, {Dohlen}, {Sauvage}, {Lee}, {Carlotti}, {Verinaud}, {Morris}, {Myers}, {Reeves}, {Swinbank}, {Calcines}, \& {Larrieu}}]{2016SPIE.9908E..1XT}
{Thatte}, N.~A., {Clarke}, F., {Bryson}, I., {et~al.} 2016, in Society of Photo-Optical Instrumentation Engineers (SPIE) Conference Series, Vol. 9908, Ground-based and Airborne Instrumentation for Astronomy VI, ed. C.~J. {Evans}, L.~{Simard}, \& H.~{Takami}, 99081X

\bibitem[{{Thomas} {et~al.}(2006){Thomas}, {Fusco}, {Tokovinin}, {Nicolle}, {Michau}, \& {Rousset}}]{2006MNRAS.371..323T}
{Thomas}, S., {Fusco}, T., {Tokovinin}, A., {et~al.} 2006, \mnras, 371, 323

\bibitem[{{V{\'e}rinaud}(2004)}]{2004OptCo.233...27V}
{V{\'e}rinaud}, C. 2004, Optics Communications, 233, 27

\end{thebibliography}

\newpage
\appendix
\section{An analytical method of computing the centroiding variance}
\label{App:CentrodingVariance}

With the matrix formalism to express the CoG algorithm, it is possible to compute the expected centroiding variances given RON and photon noise. Consider equation \ref{eq:SH_reconst}, but expressed just to compute the centroid $X$ and the centroiding error $\Xi$

\begin{equation}
    X + \Xi = M \frac{I(\phi) + b(\phi)}{N_{ph}},
\end{equation}

defining $n(\phi) = b(\phi)/N_{ph}$ we can express the centroding error as

\begin{equation}
    \Xi = M \, n(\phi)
\end{equation}

With this we can now compute the covariance matrix of the CoG:

\begin{equation}
    \label{eq:xicovar_sh}
    \left <\Xi \Xi^t \right > = M \left <n(\phi) n(\phi)^t \right > M^t,
\end{equation}

By using the statistics of the noise, we can compute the centroiding variance due to RON and photon noise. For RON, denoted as $\Sigma_{RON}$ to avoid confusion, we have that 

\begin{equation}
   \left <n(\phi) n(\phi)^t \right > = \frac{\Sigma_{RON}^2}{N_{ph}^2} \, \mathbb I,
\end{equation}

with $\mathbb I$ the identity matrix. We can replace this in the centroiding covariance matrix to obtain

\begin{equation}
    \left <\Xi \Xi^t \right > = \frac{\Sigma_{RON}^2}{N_{ph}^2} M \, M^t
\end{equation}

The diagonal of the centroiding covariance matrix would then correspond to the centroiding variance due to read-out noise. For photon noise, we have that the statistics of it follow a Poisson distribution given the light distribution on the detector. For an NGS the spot shape changes considerably in operation, but for an LGS, its shape remains approximately constant. If we consider $I_0$ to be the noiseless raw image on the detector of the SHWFS, normalized by the number of photons for the measurement to be independent of the intensity, the statistics of the noise is

\begin{equation}
    \left <n(\phi) n(\phi)^t \right > = \frac{\textbf{diag}(I_0)}{N_{ph}} \, \mathbb I,
\end{equation}

The centroiding covariance matrix for photon noise can then be expressed as

\begin{equation}
    \left <\Xi \Xi^t \right > =  \frac{1}{N_{ph}} M \, \textbf{diag}(I_0) \, M^t
    \label{appeq:cog_cov}
\end{equation}

When computing the theoretical expectations for the centroiding variance, it is generally assumed that the source is diffraction-limited or it has a Gaussian profile. For both of these cases, analytical formulas have been derived considering the statistics of photon noise. For a diffraction-limited spot, \citet{2006MNRAS.371..323T} showed that the centroiding error due to photon noise is dependent on the Field of View (FoV) of the subapertures, given by

\begin{equation}
    \sigma^2_\gamma \approx 2\frac{W}{n_{ph}},
    \label{eq:SH_DL_pn}
\end{equation}

where $W$ is the FoV of the subaperture expressed in $\lambda/d$ units, and $n_{ph}$ is the number of photons per subaperture per frame. The approximation becomes better for bigger FoV of several $\lambda/d$. For a well-sampled Gaussian profile (i.e. over 2 pixels per FWHM), with its FWHM sampled using $N_T$ pixels, and $N_{samp}$ the number of pixels used to sample a diffraction-limited spot, the formula for the centroiding error due to photon noise is

\begin{equation}
    \sigma^2_\gamma = \frac{\pi^2}{2 \ln 2} \frac{1}{n_{ph}}\left(\frac{N_T}{N_{samp}}\right)^2.
\end{equation}

\begin{figure}
    \centering
    \includegraphics[width = 0.49\textwidth]{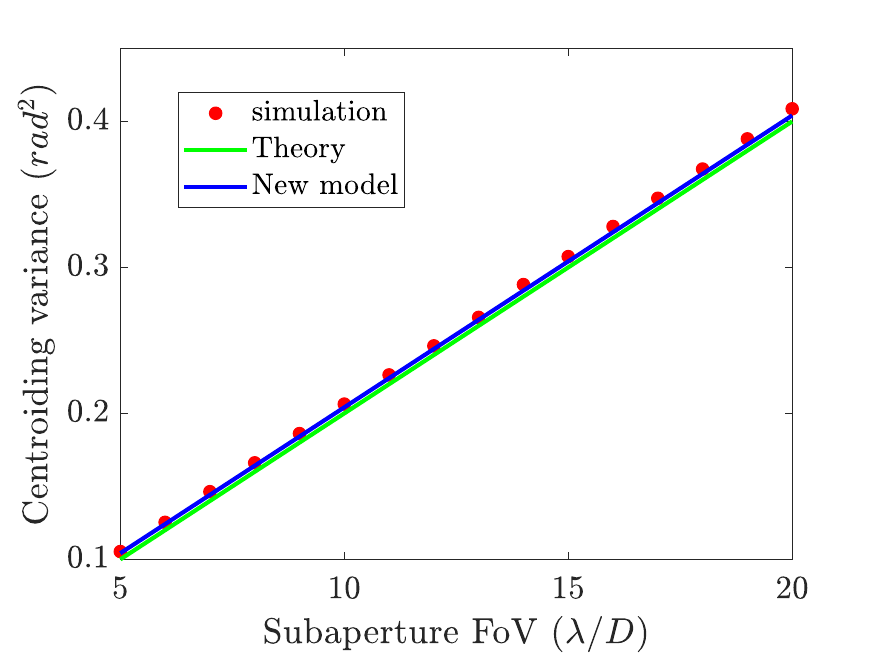}
    \caption{Centroiding error for CoG due to photon noise for a diffraction-limited source with 100 photons per subaperture per frame. The markers correspond to the E2E simulations, the green line to the theoretical expectation and the blue line to the prediction using the extended sensitivity model.}
    \label{fig:SH_DL_test}
\end{figure}

Figure \ref{fig:SH_DL_test} shows the evolution of centroiding variance due to photon noise for a diffraction-limited spot with varying FoV, which is accurately predicted by both the theoretical and the new model. The markers correspond to the centroiding variance simulation of $10^5$ independent iterations, the solid red line corresponds to the theoretical expectation using equation \ref{eq:SH_DL_pn}, considering 100 photons per subaperture and per frame, and the yellow line corresponds to the new method corresponding to the extension of the sensitivity analysis.

\begin{figure}
    \centering
    \includegraphics[width = 0.49\textwidth]{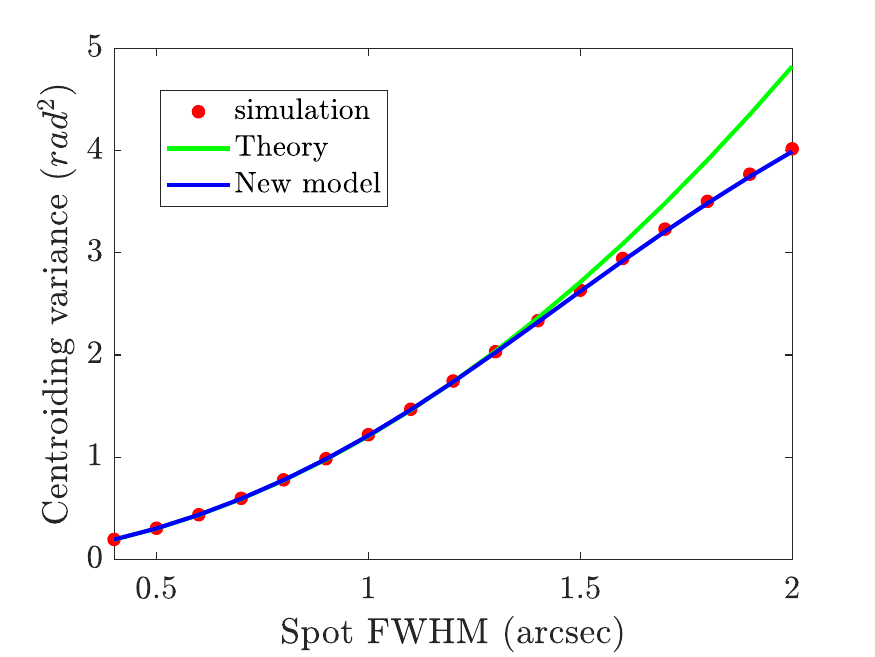}
    \caption{Centroiding error for CoG due to photon noise for a Gaussian source with 100 photons per subaperture per frame. The markers correspond to the E2E simulations, the red line to the theoretical expectation and the yellow line to the prediction using the extended sensitivity model.}
    \label{fig:SH_gauss_test}
\end{figure}

For the Gaussian spots, we set the FoV of the subaperture at 3 arcsecs and then varied the FWHM of the spot and did the same number of simulations as with the diffraction-limited case. Figure \ref{fig:SH_gauss_test} shows the results from the E2E simulations, theory and the new model, again with good agreement between them. We intentionally set the FoV of the subaperture comparable to the biggest FWHM tested, to emphasize the fact that the theoretical formula assumes that the FoV is much bigger than the spot, but in reality, this spot gets cropped, an effect that is not taken into account by the theoretical formula, but is better predicted by this new model.

\begin{figure}
    \centering
    \includegraphics[width = 0.49\textwidth]{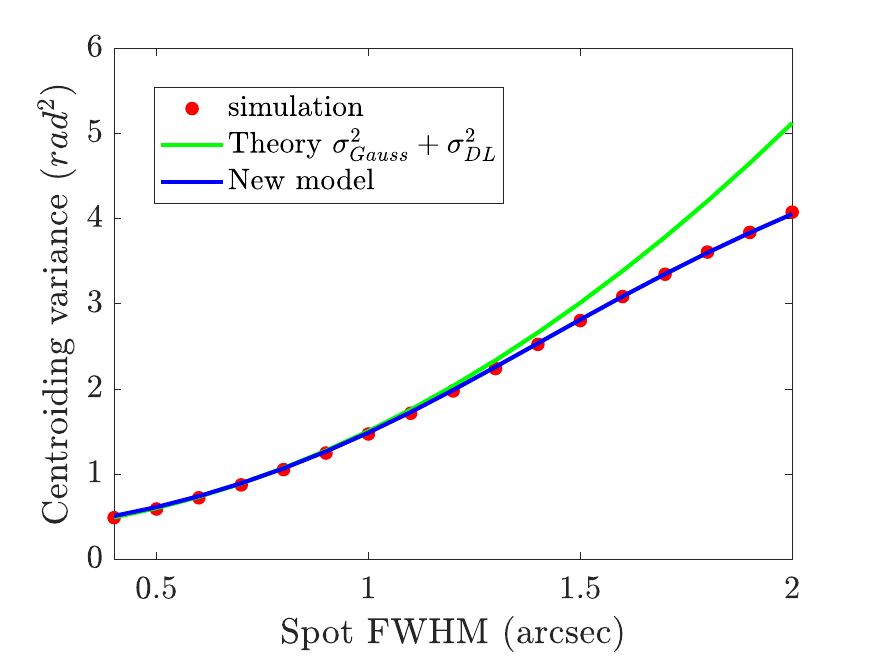}
    \caption{Centroiding error for CoG due to photon noise for the convolution of a Gaussian source with the diffraction-limited spot with 100 photons per subaperture per frame, a field of view of $15 
    \, \lambda/D$ and a pixel scale of $0.12 \, arcsec/pix$. The markers correspond to the E2E simulations, the red line to the theoretical expectation and the yellow line to the prediction using the extended sensitivity model.}
    \label{fig:SH_DL_conv_gauss_test}
\end{figure}

Theoretical predictions of the centroiding variance cannot reach much further than these two cases: there are no closed formulas for when the light distribution is not one of these cases. In appendix \ref{app:CentroidingConvolvedImages}, we show that it is possible to express the CoG variance of the convolution of two sources (e.g. the convolution of the diffraction-limited spot and a Gaussian object), as the sum of the centroiding variances of each separately. 
To test this, we simulated using E2E propagation the spot for a Gaussian object, that was affected by the size of the optics, meaning that it was convolved with the diffraction-limited spot. We then varied the FWHM of the Gaussian object and recorded the evolution of the centroiding error due to photon noise, as can be observed in Fig. \ref{fig:SH_DL_conv_gauss_test}. Both the theoretical predictions and the new model can predict the behavior of the simulation. 

\begin{figure}
    \centering
    \includegraphics[width = 0.49\textwidth]{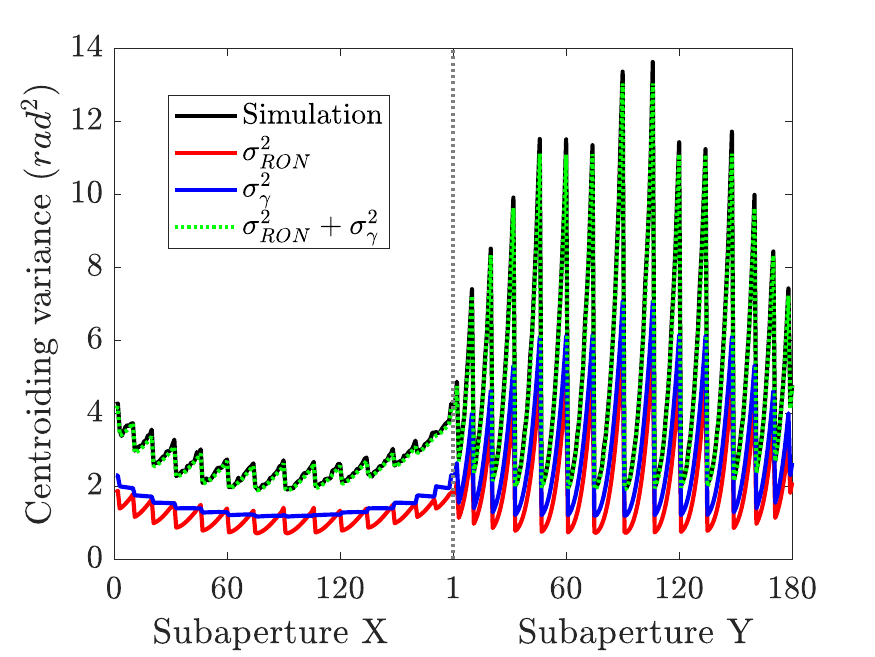}
    \caption{Centroiding error for WCoG due to RON and photon noise for an LGS with an 8 m telescope and a 16 x 16 subaperture SH with 100 photons per subaperture per frame. The black curve corresponds to the E2E simulations, the red line to the predicted centroiding variance due to RON, the blue line to the predicted variance due to photon noise and the green line the sum of the two. Both X and Y centroiding variances are shown in the same plot. the left part of the plot is for the variance along the X axis and the right part of the plot is the variance along the Y axis}
    \label{fig:LGS_Centroiding}
\end{figure}

We have shown that the new method using equation \ref{appeq:cog_cov} can replicate the results predicted by the theory. The benefit of this new method is that it can predict the centroiding variance for any shape of the spot. This is particularly useful for its use with LGS, as each subapertures has a different spot shape, and depending on the sodium density profile, the spots themselves can differ from Gaussian objects. Also, as the orientation of the spot depends on the physical position of the subaperture, we don't need to assume that the object's elongation is along one of the axes. Finally, this model allows to predict the centroiding variance when using WCoG, a task that can be too complicated to do in theory. Fig. \ref{fig:LGS_Centroiding} shows a simulation of 10, 000 independent iterations of centroiding for a 16 x 16 subaperture SH with an LGS for an 8 m telescope. We added RON of $2 \, e^{-}/pix/frame$ and computed the expected centroiding variance for both RON and photon noise. The total expected centroiding variance accurately predicted the one obtained in the independent simulation. We plotted both the centroiding in the X and Y direction in the plot, and there is a clear difference in behavior between the two, given that the elongation of the LGS is along the Y axis. 

\section{Centroiding variance for convolved images}
\label{app:CentroidingConvolvedImages}
An image $P(x,y)$ in the detector can be expressed as the convolution of two images $f(x, y)$ and $g(x, y)$ (e.g. diffraction limited and Gaussian spots)

\begin{equation}
    P(x,y) = n_{ph}\,f(x,y) \ast g(x,y),
\end{equation}

with $n_{ph}$ the number of photons in the image. Here the images have to be normalized such that

\begin{equation}
    \iint f(x,y) dx dy = \iint g(x,y) dx dy = 1.
\end{equation}

Consider the following expressions to compute the CoG and the centroiding variance due to photon noise

\begin{equation}
\begin{split}
\overline x_P  & = \frac{\iint x \, P(x,y) dx\,dy}{\iint P(x,y) dx\,dy} = \iint x \, f(x,y) \ast g(x,y) dx\,dy \\
\sigma_{x_P}^2 & = \frac{\iint (x - \overline x_P)^2 \, P(x,y) dx\,dy}{\iint P(x,y) dx\,dy} \\ & = \iint (x - \overline x_P)^2 \, f(x,y) \ast g(x,y) dx\,dy.
\end{split}
\label{eq:CoGandVariance}
\end{equation}

Let us consider both $f$ and $g$ as probability density functions (PDF). If we sum the two PDF, the resulting PDF is obtained by the convolution of $f$ and $g$, which is what we are interested in. The resulting PDF will have a mean (or CoG) equal to the sum of the means of each independent PDF, and the variance of the new PDF will be equal to the sum of the variances of $f$ and $g$. Therefore, considering the expressions for the CoG and the centroiding variances given in \ref{eq:CoGandVariance}, for the convolution of two images we have

\begin{equation}
    \begin{split}
        \overline x_P & = \overline x_f + \overline x_g \\
        \sigma_{x_P}^2 & = \sigma_{x_f}^2 + \sigma_{x_g}^2.
    \end{split}
\end{equation}

This derivation is not possible is using WCoG, as then cross-correlation terms appear in the expression that do not allow to separate the contribution to the centroiding variance of each image that is convolved.

\section{Estimating the centroiding variance for multiple LGS systems}
\label{App:multipleLGS}

If we have knowledge about the centroiding covariance matrix, it is possible to combine the measurements of multiple LGS with a Generalized least squares estimator. If we have $N$ LGSs, for each subaperture we have $N$ measurements $x_1', x_2', \dots, x_N'$ and N measurements $y_1', y_2', \dots, y_N'$. Given the direction of the elongation of the LGS, $x_i'$ and $y_i'$ might be correlated, and instead of measuring along the regular x/y axis, a rotated axis might be used to align the measurements with the elongation direction. If we combine all the measurements into a single linear system, in matrix notation we have

\begin{equation}
    \begin{pmatrix}
        x_1' \\ x_2' \\ \vdots \\ x_N' \\ y_1' \\ y_2' \\ \vdots \\ y_N'
    \end{pmatrix} 
    =
    \begin{pmatrix}
        cos(\theta_1) & sin(\theta_1) \\
        cos(\theta_2) & sin(\theta_2) \\
        \vdots \\
        cos(\theta_N) & sin(\theta_N) \\
        -\sin(\theta_1) & cos(\theta_1) \\
        -\sin(\theta_2) & cos(\theta_2) \\
        \vdots \\
        -\sin(\theta_N) & cos(\theta_N) \\
    \end{pmatrix}
    \begin{pmatrix}
        x \\
        y
    \end{pmatrix},
\end{equation}

with $\theta_i$ the rotated angle of the axis in which the measurements are made, and $x,y$ are the unknown centroid coordinates. In matrix form, the equation can be written as

\begin{equation}
    \bold x' = B \bold x.
\end{equation}

If we have the centroiding covariance matrix $\Sigma$ of the $N$ $x_i'$ $y_i'$ measurements, the solution to the Generalized least squares problem is

\begin{equation}
    \bold x = (B^T \Sigma^{-1} B)^{-1} B^T \Sigma^{-1} \bold x',
\end{equation}

from which the covariance of the estimates $\bold x$ is

\begin{equation}
    Cov(\bold x) = (B^T \Sigma^{-1} B)^{-1}.
\end{equation}

\begin{figure}
    \centering
    \includegraphics[width=0.49\textwidth]{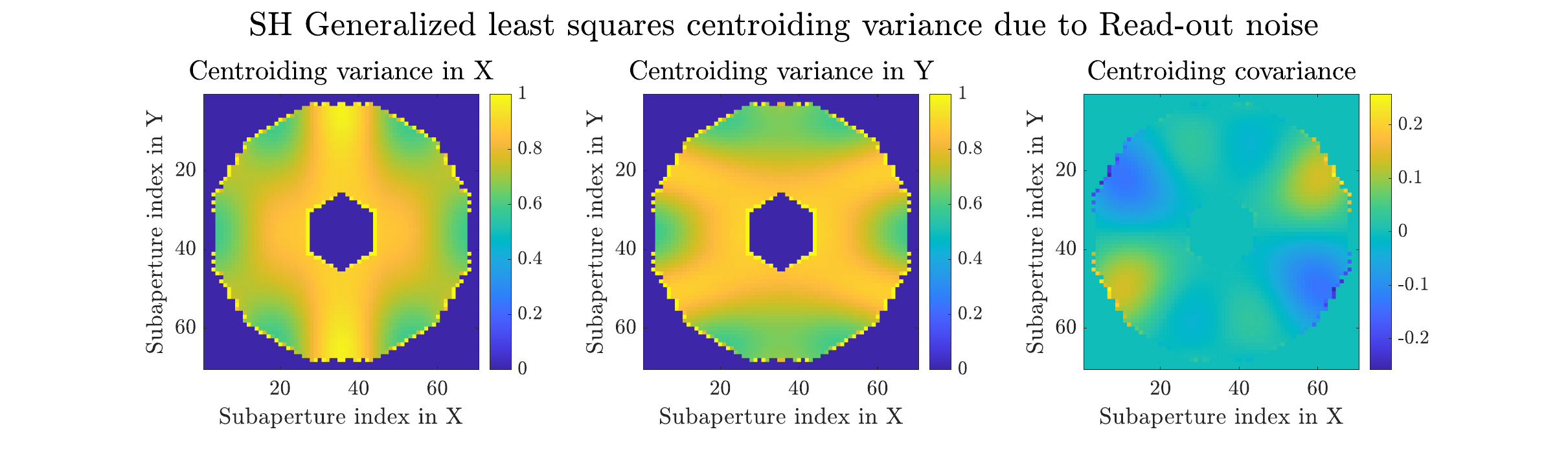}\\
    \includegraphics[width=0.49\textwidth]{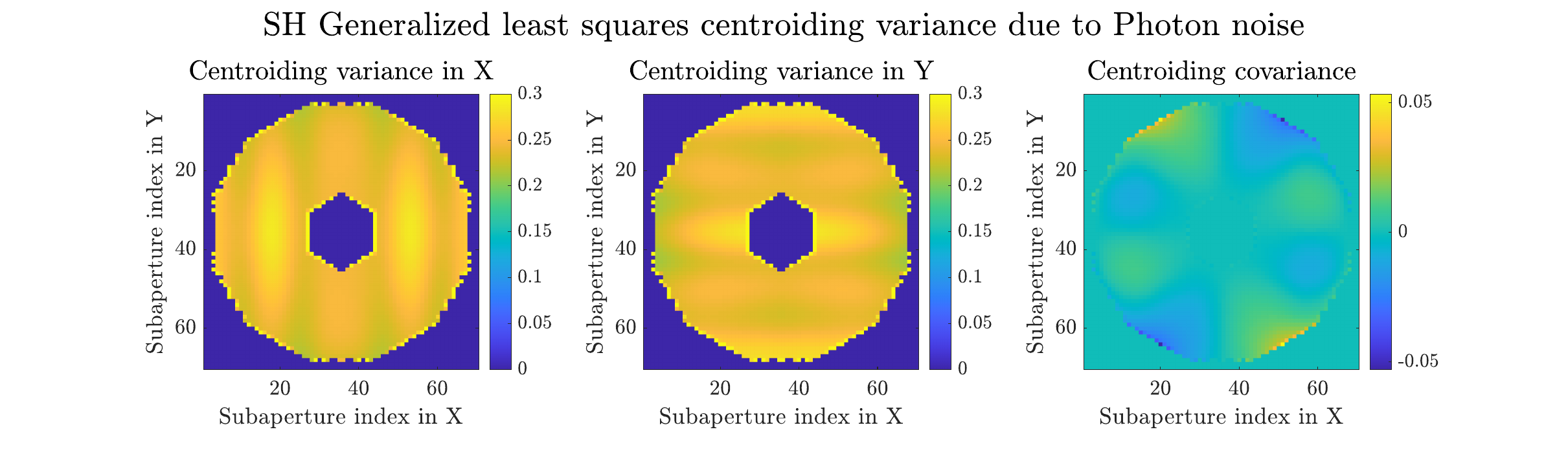}
    \caption{SH centroiding variances in X and Y direction for a 40 m telescope resulting from the combination of 6 LGSs with the Generalized least square. The variances are normalized by the variance of a 1 arcsec spot.}
    \label{fig:GLSCentroidingVarianceSH}
\end{figure}

In this work, we use 6 LGSs around the primary mirror. Figure \ref{fig:GLSCentroidingVarianceSH} shows the centroiding variance in the X and Y direction at each subaperture, resulting from the combination of the measurements from each LGS. The variances are normalized by the variance of a 1 arcsec spot. For RON, the average centroiding variance in both X and Y directions is 0.81 times the centroiding variance of a 1 arcsecond spot, meanwhile, for photon noise it corresponds to 0.24 times the variance of a 1 arcsecond spot.

\begin{figure}
    \centering
    \includegraphics[width=0.49\textwidth]{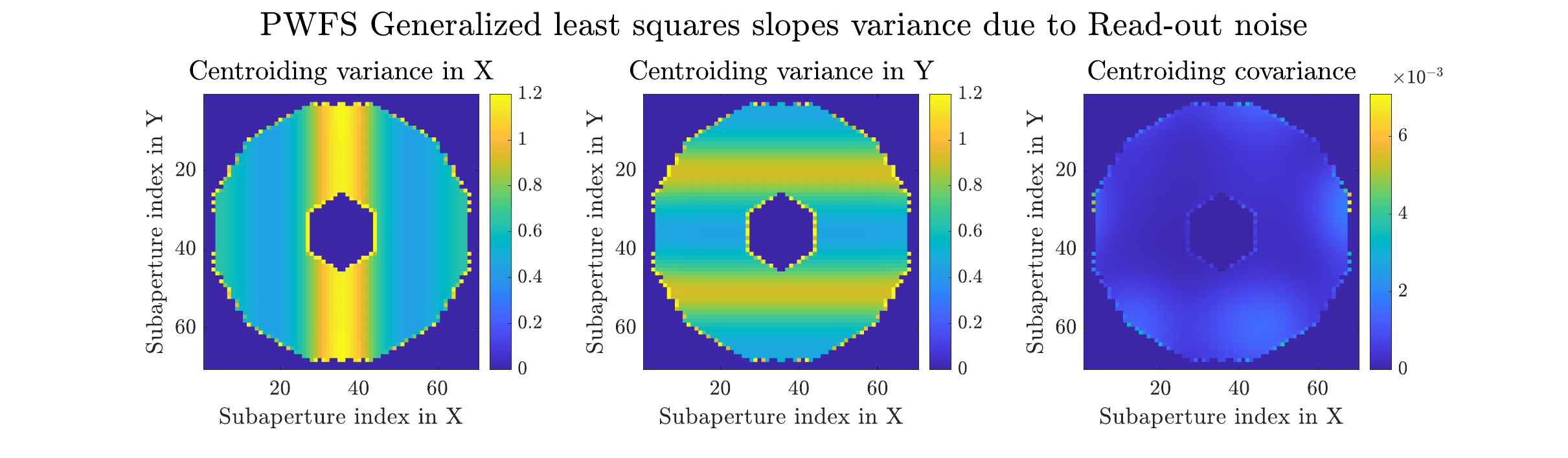}\\
    \includegraphics[width=0.49\textwidth]{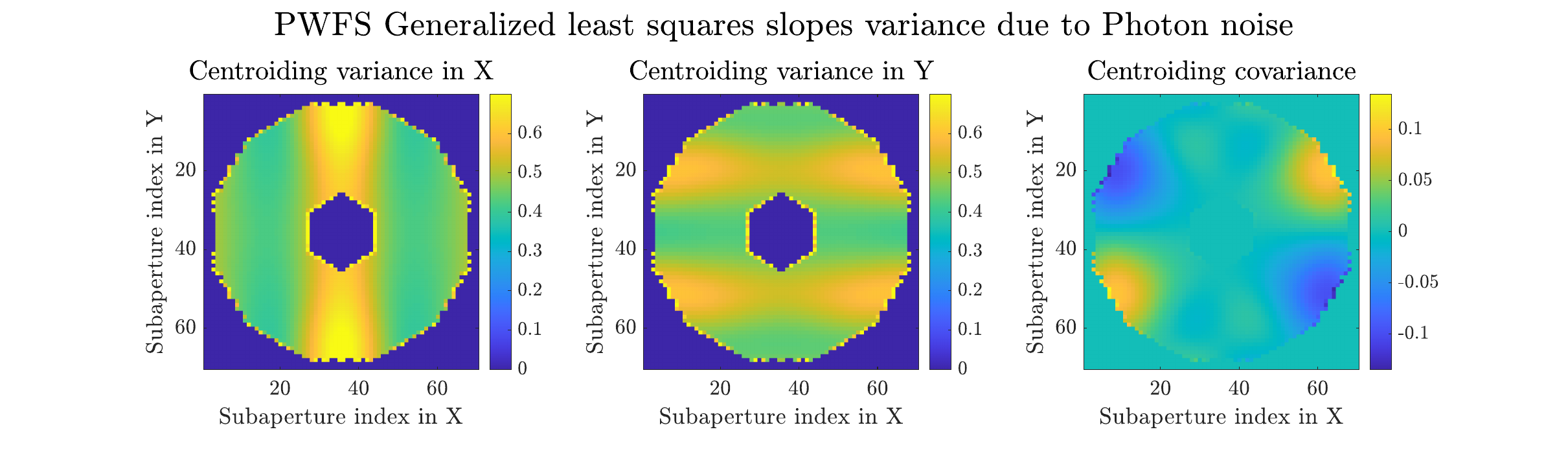}
    \caption{PWFS variances in X and Y direction for a 40 m telescope resulting from the combination of 6 LGSs with the Generalized least square. The variances are normalized by the variance of a 1 arcsec spot. 16 x 16 subapertures are used here for visualization purposes.}
    \label{fig:GLSCentroidingVariancePWFS}
\end{figure}

For the PWFS, it is possible to follow the same formalism as the SH shown in App. \ref{appeq:cog_cov} to obtain the slopes-X and slopes-Y covariance matrix for RON and photon noise. By using the Generalized least square, it is possible to obtain the variance of the combined slopes-X and slopes-Y of the 6 LGSs, as shown in figure \ref{fig:GLSCentroidingVariancePWFS}. The variances are normalized by the variance of a 1-arcsecond spot. For RON, the average variance is 0.70 times that of the 1-arcsecond spot and for photon noise is 0.50 times that of the 1-arcsecond spot.

\end{document}